\documentclass[twocolumn,showpacs,showkeys,preprintnumbers,superscriptaddress,amsmath,floatfix,amssymb,secnumarabic,nofootinbib]{revtex4-1}
\usepackage[colorlinks=true]{hyperref}
\usepackage{graphicx}
\usepackage{dblfloatfix}    
\usepackage{placeins}
\usepackage{braket}
\usepackage{amsmath}
\usepackage{epsfig}
\usepackage{float}
\usepackage{nicefrac}
\usepackage{xcolor}
\usepackage{color}
\usepackage{bm}
\usepackage{subfigure}

\def\({\left(}
\def\){\right)}
\def\[{\left[}
\def\]{\right]}

\newcommand{\vev}[1]{ \langle \, #1 \, \rangle }
\newcommand{\diag}[1]{ {\rm diag} \, \left( #1 \right) }
\newcommand{\Tr}{ {\rm Tr} \, }

\newcommand{\beq} {\begin{eqnarray}}
\newcommand{\eeq} {\end{eqnarray}}

\newcommand{\comment}[1]{}

\begin{document}
\sloppy

\title{Lefschetz thimbles decomposition for the Hubbard model on the hexagonal lattice}

\author{Maksim~Ulybyshev}
\email{Maksim.Ulybyshev@physik.uni-wuerzburg.de}
\affiliation{Institute for Theoretical Physics, Julius-Maximilians-Universit\"at W\"urzburg,
   97074 W\"urzburg, Germany}

\author{Christopher Winterowd}
\email{c.r.winterowd@kent.ac.uk}
\affiliation{University of Kent, School of Physical Sciences, Canterbury CT2 7NH, UK}

\author{Savvas Zafeiropoulos}
\email{savvas.zafeiropoulos@cpt.univ-mrs.fr}
\affiliation{Institute for Theoretical Physics, Heidelberg University, Philosophenweg 12, 69120 Heidelberg, Germany}
\affiliation{Aix Marseille Univ, Universit\'e de Toulon, CNRS, CPT, Marseille, France}

\begin{abstract}
In this article we propose a framework for studying the properties of the Lefschetz thimbles decomposition for lattice fermion models approaching the thermodynamic limit. The proposed set of algorithms includes the Schur complement solver and the exact computation of the derivatives of the fermion determinant. It allows us to solve the gradient flow (GF) equations taking into account the fermion determinant exactly, with high performance. Being able to do so, we can find both real and complex saddle points and describe the structure of the Lefschetz thimbles decomposition for large enough lattices which allows us to extrapolate our results to the thermodynamic limit. We describe the algorithms for a general lattice fermion model, with emphasis on two widely-used types of lattice discretizations for relativistic fermions(staggered and Wilson fermions), as well as on interacting tight-binding models for condensed matter systems. As an example, we apply these algorithms to the Hubbard model on a hexagonal lattice. Several technical improvements allow us to deal with lattice volumes as large as $12 \times 12$ with $N_\tau=256$ steps in Euclidean time, in order to capture the properties of the thimbles decomposition as the thermodynamic, low-temperature, and continuum limits are approached.  Different versions of the Hubbard-Stratonovich (HS) transformation were studied, and we show that the complexity of the thimbles decomposition is very dependent on its specific form. In particular, we provide evidence for the existence of an optimal regime for the hexagonal lattice Hubbard model, with a reduced number of thimbles becoming important in the overall sum.  In order to check these findings, we have performed quantum Monte Carlo (QMC) simulations using the gradient flow to deform the integration contour into the complex plane. These calculations were made on small volumes ($N_s=8$ sites in space), albeit still at low temperatures and with the chemical potential tuned to the van Hove singularity, thus entering into a regime where standard QMC techniques exhibit an exponential decay of the average sign. The results are compared versus exact diagonalization (ED), and we demonstrate the importance of choosing an optimal form for the HS transformation for the Hubbard model to avoid issues associated with ergodicity. We compare the residual sign problem with the state-of-the-art BSS (Blankenbecler, Scalapino and Sugar)-QMC and show that the average sign can be kept substantially higher using the Lefschetz thimbles approach. 
\end{abstract}

\maketitle

\section{\label{sec:Intro}Introduction}
One of the most widely used paths to study, non-perturbatively, the physics of strongly coupled quantum systems, in a fully \textit{ab-initio} manner is through Monte Carlo simulations of the Feynman path integral. For many systems, the Euclidean formulation of the functional integral yields a real, positive-definite action whose evaluation can be performed using importance sampling. It often occurs, however, that the action is complex. Relevant examples of such systems abound in disparate branches of physics.  The theory of the strong interactions, quantum chromodynamics (QCD), exhibits a complex action at finite baryon density~\cite{Philipsen:2010gj,Aarts:2015tyj}. The study of the QCD phase diagram is essential to the understanding of the quark-gluon plasma, a strongly-coupled fluid which exists above the deconfinement transition, as well addressing deep questions in astrophysics and cosmology. Currently, these phases of matter are actively studied experimentally at several collider facilities, including RHIC and LHC.
Additional examples from high-energy physics which are well-known to exhibit a complex action include gauge theories with the addition of a theta term, Chern-Simons theories, and matrix models which provide a non-perturbative definition of string theories. 
The sign problem also impedes \textit{ab-initio} studies of many-body systems. The physics of Feshbach resonances in cold atomic Fermi gases, frequently realized in the laboratory and modeled with the unitary Fermi gas (UFG) is one prominent example~\cite{Inguscio:2007cma,Schafer:2009dj,Rammelmuller:2018hnk}. A well-studied  example from condensed matter physics is the Hubbard model. Despite its simplicity, the Hubbard model captures the physics of the Mott metal-insulator transition, and probably, high-temperature superconductors~\cite{Baeriswyl,Lee:2006zzc}. On a bipartite lattice at half-filling, the Hubbard model is free from the sign problem due to particle-hole symmetry. However, as soon as frustration or non-zero chemical potential appear in the Hamiltonian, one is faced with the sign problem.  In fact, it could be argued that the majority of interesting systems are plagued by the sign problem. A key result, due to Troyer and Wiese~\cite{Troyer:2004ge}, states that the sign problem is an NP-hard problem in a generic, Ising spin-glass system. It follows that a general solution to all sign problems is an unlikely proposition. 

There are several approaches one can take to dealing with the problem. The most naive approach to deal with a system exhibiting a complex action is to absorb the imaginary part of the action in the observable and sample according to the real part of the action.
This method is known as reweighting and is based on the following identity
\begin{align}
\label{eq:reweighting_identity}
    &\langle\mathcal{O}\rangle = \frac{1}{\mathcal{Z}} \int\mathcal{D}\Phi\,\mathcal{O}[\Phi]\,e^{-S[\Phi]} = \frac{\int\mathcal{D}\Phi\,\mathcal{O}[\Phi]\,e^{-S[\Phi]}}{\int\mathcal{D}\Phi\,e^{-S[\Phi]}}&\nonumber\\
    &= \frac{\frac{1}{\mathcal{Z}_{\mathrm{pq}}}\int\mathcal{D}\Phi\,\mathcal{O}[\Phi]\,\frac{e^{-S[\Phi]}}{e^{-S_R[\Phi]}}\,e^{-S_R[\Phi]}}{\frac{1}{\mathcal{Z}_{\mathrm{pq}}}\int\mathcal{D}\Phi\,\frac{e^{-S[\Phi]}}{e^{-S_R[\Phi]}}\,e^{-S_R[\Phi]}}
    =\frac{\langle\mathcal{O}e^{-iS_I}\rangle_{S_R}}{\langle e^{-iS_I}\rangle_{S_R}},&
\end{align}
where $S=S_R + i S_I$, the ratio $\nicefrac{e^{-S[\Phi]}}{e^{-S_R[\Phi]}}$ is the reweighting factor, and
\begin{equation}
\label{eq:phase_quenched_partition_function}
\mathcal{Z}_{\mathrm{pq}}=\int\mathcal{D}\Phi\,e^{-S_R[\Phi]} 
\end{equation}
is the phase quenched partition function.
The angular brackets in (\ref{eq:reweighting_identity}) denote an ensemble average with respect to the measure  $\mathcal{D}\Phi\;e^{-S_R}$.
Although this sequence of expressions is nothing more than a rewriting of the standard thermal ensemble average, the practical calculation of observables using reweighting is exponentially difficult due to the sign problem. The last ratio in (\ref{eq:reweighting_identity}) is not well defined, since both the numerator and the denominator vanish exponentially as the spacetime volume is increased. This is a manifestation of what is coined as the overlap problem, i.e.\ the phase-quenched theory is different from the full theory which renders sampling from the former a highly ineffective approach to mimic sampling from the full theory. 
The technical issue at hand is the overlap of the ensemble sampled according to $S_R$  and the original ensemble that involves the entire action.
A physical meaning can be attached to this difficulty by considering the average sign, $\langle e^{-iS_I}\rangle_{S_R}$, which can be understood as a ratio of two partition functions
\begin{equation}
\label{eq:ratio_partion_functions} 
    \frac{\mathcal{Z}}{\mathcal{Z}_{\mathrm{pq}}} = e^{- \beta V \Delta f}.
\end{equation}
In (\ref{eq:ratio_partion_functions}) we have introduced the spatial volume $V$, inverse temperature $\beta$, and  $\Delta f$, which is the free energy density difference between the two ensembles. Although $\Delta f$ is formulation dependent, one cannot cure the exponential scaling using naive reweighting. In any Monte Carlo calculation, the error on the mean scales with the computational time, $T_{CPU}$, as $1/\sqrt{T_{CPU}}$. Thus, in order to have the error on the average sign less than the value of the average sign itself, we must require that $T_{CPU} \gg e^{2\beta V \Delta f}$. We refer the interested reader to~\cite{Aarts:2015tyj} for a pedagogical and detailed presentation of the sign and overlap problems.

Recently, much progress has been made by complexifying the fields of systems suffering from the sign problem.
This idea, which can easily be demonstrated in simple, one-dimensional integrals, has already been applied to a number of models, which we list below. Furthermore, in many of these cases,  the initial severe sign problem was eliminated or substantially weakened. 
One successful approach along the previously mentioned lines is complex Langevin dynamics\cite{Aarts:2013lcm,Aarts:2013uxa,Sexty:2013ica,Nagata:2016alq,Aarts:2017vrv,Bloch:2017sex,Anagnostopoulos:2017gos,Scherzer:2018hid}. Another method, which we use in this study is the method of Lefschetz  thimbles. Originally introduced in~\cite{Witten:2010zr,Witten:2010cx}, it was not long after that lattice gauge theory practitioners sought to apply these methods to QCD at finite baryon density~\cite{Cristoforetti:2012su}. Pioneering studies using Lefschetz thimbles were performed on the relativistic Bose gas for lattices volumes up to $V=8^4$, showing good agreement with complex Langevin simulations~\cite{Cristoforetti:2013wha,Fujii:2013sra,Cristoforetti:2013qaa}. Several other studies have investigated a variety of other systems displaying a sign problem including $O(n)$ sigma models~\cite{Tanizaki:2014tua}, chiral random matrix ensembles \cite{DiRenzo:2015foa}, and the $U(1)$ one-link model using techniques borrowed from reweighting~\cite{Bluecher:2018sgj}. A significant hurdle was overcome as several groups extended the method of Lefschetz thimbles to interacting fermions in $0+1$ dimensions as well as at a single site~\cite{Fujii:2015bua,Tanizaki:2015rda,Kanazawa:2014qma,Alexandru:2015xva,Alexandru:2015sua,DiRenzo:2017igr}. The successful application of these methods was then subsequently extended to field theories of strongly interacting fermions in both $1+1$ as well as $2+1$ dimensions~\cite{Alexandru:2016ejd,Alexandru:2018brw,Alexandru:2018ngw,Alexandru:2018ddf}.  
A short description of the results presented in this article and related to the Hubbard model originally appeared in our previous letter~\cite{Ulybyshev:2019hfm}. In a subsequent preprint ~\cite{Fukuma:2019wbv}, an alternative approach to deal with ergodicity issues was applied to Hubbard model simulations within the thimbles formalism. Albeit, there is a big difference in the regimes studied by the two groups. In~\cite{Ulybyshev:2019hfm} and in this paper, the low temperature limit of the Hubbard model was studied, while the authors of~\cite{Fukuma:2019wbv} reported results for significantly higher temperatures and thus of milder sign problem for which BSS-QMC has an average sign greater than 0.6 in the whole range of parameters studied in their paper. On the other hand, both in~\cite{Ulybyshev:2019hfm} as well as in this article, we are addressing the region of strong sign problem where BSS-QMC, even with an optimal setup, experiences an exponential decay of the average sign.

One of the main concerns about the efficacy of the Lefschetz thimbles method is the scaling of the number of thimbles with the system size and temperature. One might argue that the amount of important thimbles can grow exponentially once we approach the thermodynamic or low temperature limit. In this case, by construction, the method is unable to improve the sign problem. The problem is especially hard for fermionic models, since it is usually quite difficult to find the exact saddle points if the logarithm of the fermionic determinant is included in the action.  Here, we propose a set of algorithms to address this issue and to elucidate, clearly and systematically, the non-trivial saddle point structure of the theory. Due to a more efficient calculation of the exact derivatives of the fermionic determinant, we are now able to reveal the construction of the Lefschetz thimble decomposition on large lattices and extrapolate our results to the thermodynamic limit. This also represents the main difference of our paper from earlier attempts to apply the Lefschetz thimbles decomposition to the Hubbard model \cite{PhysRevB.90.035134}, where the thimbles decomposition was not optimised and only one thimble, out of many important ones, was taken into account. As a result, those simulations actually did not represent a full calculation of the functional integral, but rather represented only corrections to dynamical mean field theory (DMFT) results. Using a complete study of the saddle point structure of the Hubbard model, and identifying the advantageous regions in parameter space, one can safely proceed to address the sign problem using Lefschetz thimbles.

We start with a short introduction to the formalism, and proceed with the description of the method to solve the gradient flow equations for Wilson and staggered fermions. After this, we describe the application of the technique to the Hubbard model on the hexagonal lattice. First, we make a detailed study of the saddle points, which is an essential ingredient of the Lefschetz thimbles method. In particular, we explore the dependence of saddles on volume, the Hubbard coupling $U$, and chemical potential. Among other things, we discuss at length the algorithms used to search for saddle point configurations away from half-filling, when saddle points are shifted into complex space $\mathbb{C}^N$. Finally, in order to support our conclusions concerning the role of different saddle points, we perform Monte Carlo calculations over manifolds in complex space and compare results with exact diagonalization. In addition to that, we show that the average sign can be substantially increased even in comparison with BSS-QMC. This fact means that we can potentially construct a superior algorithm for dealing with the sign problem, if the additional computational costs associated with the gradient flow and integration over curved manifolds in complex space are improved upon. 

\section{\label{sec:formalism}Lefschetz thimbles formalism}
Let us first consider the complexification of the fields appearing in the functional integral (\ref{eq:reweighting_identity}), $\Phi \in \mathbb{C}^N$. This amounts to a shift of the contour of integration into complex space. We are allowed to do so, as Cauchy's theorem tells us that one can choose any appropriate contour in complex space as long as the integral still converges and no poles of the integrand are crossed during this shift. As we will demonstrate, both of these conditions are satisfied. We now introduce one particularly useful representation, known as the Lefschetz thimble decomposition of the partition function \cite{Witten:2010zr,Witten:2010cx},
\begin{align}
\label{eq:thimbles_sum_and_integral}
&\;\;\mathcal{Z} = \int_{\mathbb{R}^N} \mathcal{D} \Phi\, e^{-S[\Phi]}=\sum_\sigma k_\sigma \mathcal{Z_\sigma},&\nonumber\\
\text{where}\quad&\mathcal{Z_\sigma} = \int_{\mathcal{I}_\sigma} \mathcal{D} \Phi\, e^{-S[\Phi]},&
\end{align}
and $\sigma$ labels all complex saddle points $z_\sigma \in \mathbb{C}^N$ of the action, which are determined by the condition 
\begin{equation}
  {\left.\frac{\partial S}{\partial \Phi}\right| }_{\Phi=z_\sigma} = 0.  
\end{equation}
The integer-valued coefficients $k_\sigma$, are the intersection numbers and $\mathcal{I_\sigma}$ are the Lefschetz thimble manifolds attached to the saddle points $z_\sigma$. These manifolds, defined below, are the generalization of the contours of steepest descent in the theory of asymptotic expansions. We stress that if the saddle points are non-degenerate (${\left.\det \partial^2 S/\partial \Phi' \partial \Phi\right| }_{\Phi=z_\sigma} \neq 0$) and isolated, the relation (\ref{eq:thimbles_sum_and_integral}) holds (for a generalization to the case of gauge theory see \cite{Witten:2010cx}).

\begin{figure}
        \centering
        \includegraphics[scale=0.95, angle=0]{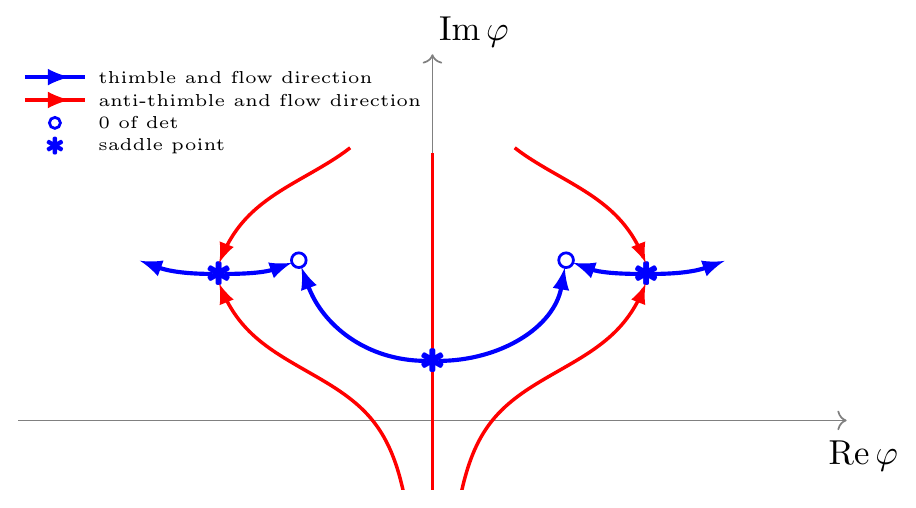}
        \caption{Typical scheme of thimbles ant anti-thimbles, arrows show the directions of the flows, which define these manifolds.}
        \label{fig:initial_thimbles_scheme}
\end{figure}

The Lefschetz thimble manifold associated with a given saddle point is the union of all solutions of the following differential equation
\begin{equation}
\label{eq:flow}
\frac{d\Phi}{d t}=\overline{ \frac{\partial S}{\partial \Phi}},
\end{equation}
known as the gradient flow (GF) equations, which satisfy the following boundary condition: $~\Phi\in\mathcal{I}_\sigma: \Phi( t \rightarrow -\infty) \rightarrow z_\sigma$. Just as we made an analogy between the thimble and the contour of steepest descent, there is a second manifold associated with each saddle point which is analogous to the contour of steepest ascent. This manifold is known as the anti-thimble, $\mathcal{K}_\sigma$, and consists of all possible solutions of the GF equations (\ref{eq:flow}) which end up at a given saddle point $z_\sigma$:  $\Phi\in\mathcal{K}_\sigma:\Phi(t )=\Phi, \Phi(t \rightarrow + \infty) \rightarrow z_\sigma$.  The intersection number $k_\sigma$ is defined by counting the number of intersections of  $\mathcal{K}_\sigma$ with the original integration domain: $\mathbb{R}^N$, $k_\sigma = \langle \mathcal{K}_\sigma, \mathbb{R}^N \rangle$. An example scheme of thimbles and anti-thimbles is drawn in the Fig.~\ref{fig:initial_thimbles_scheme}.

It is worth noting that thimbles and anti-thimbles are both real, $N$-dimensional manifolds embedded in $\mathbb{C}^N$. We now state two key properties of the thimbles, which follow from (\ref{eq:flow}) coupled with the fact that the action, $S$, is regarded as a holomorphic function of the complex fields. These properties are that the real part of the action, $\mbox{Re}\, S$, monotonically increases along the thimble, starting from the saddle point and the imaginary part of the action, $\mbox{Im}\, S$, stays constant along it. The first property is essential in guaranteeing the convergence of the individual integrals in (\ref{eq:thimbles_sum_and_integral}), while the second one obviously makes the method attractive with regards to the weakening of the sign problem. 
Using these crucial properties, it follows that neither thimbles nor anti-thimbles can intersect each other, no two saddle points can, in general, be connected by a thimble (with the very important exception which is discussed below), and all integrals on the r.h.s. of (\ref{eq:thimbles_sum_and_integral}) are convergent.

As a result of the above discussion, it follows that (\ref{eq:thimbles_sum_and_integral}) can be written as
\begin{eqnarray}
\label{eq:thimbles_sum_with_phases}
\mathcal{Z} = \sum_\sigma k_\sigma e^{-i\, \mathrm{Im}\, S(z_\sigma)} \int_{\mathcal{I}_\sigma} \mathcal{D} \Phi\, e^{-\mathrm{Re} \, S(\Phi)},
\end{eqnarray}
where we have explicitly written out the complex factors associated with different saddle points. 
Usually, thimbles can be classified as being either ``relevant" or ``irrelevant" using the intersection number.  Relevant ones have their intersection number, $k_\sigma$, being nonzero and thus participate in the sum in (\ref{eq:thimbles_sum_with_phases}). Conversely, a thimble is irrelevant if it has a zero intersection number. However, this classification can fail if the so-called Stokes phenomenon occurs for saddle points lying within $\mathbb{R}^N$. By definition, the Stokes phenomenon means that the saddle points are connected by a thimble. In this case, $k_\sigma$ is not well-defined and we need other tools in order to classify the saddle points. An example of such a situation will be demonstrated later on when we will study the actual saddle points for the Hubbard model. 

As one can see, the initial sign problem has been split into two parts. The first part of the residual sign problem concerns the constant phase factors, $e^{-i\, \mathrm{Im}\, S(z_\sigma)}$. The number of relevant thimbles, their weight, and the distribution of the imaginary part of the action at corresponding saddles define the remaining severity of the first part of the sign problem. An ideal situation arises when this sum only contains one dominant term. The second part of the residual sign problem relates to the fluctuations of the complex measure, $\mathcal{D} \Phi$, in the integration over the thimble. Potentially, there is the third source of the residual sign problem: residual fluctuations of $\mathrm{Im} \,S$, which appear if we are not following the thimble exactly. All these issues will be addressed below in our test calculations for the Hubbard model, where we perform a thorough classification of saddle points and then give an estimate for the fluctuations of the complex measure and the residual fluctuations of  $\mathrm{Im} \,S$. We now present a description of our numerical methods.

\section{\label{sec:algoriths}Algorithms}

The GF equations (\ref{eq:flow}) are the basis of the whole formalism. Here we present the set of algorithms, which allows us to solve them efficiently at least for lattices of moderate size. The main difficulty in implementing GF is the presence of the fermionic determinant in the action for a typical lattice field theory (or model -for the case of condensed matter systems) with fermions
\begin{equation}
\label{eq:S_eff}
S=S_b +\ln \det M,
\end{equation}
where $S_b$ is the bosonic part and the fermionic operator $M$ is more or less a sparse matrix with dimensionality  $\sim N_\tau N_s$ (ignoring for the moment color and flavor indices). Here, $N_\tau$ is the Euclidean time extent of the lattice and $N_s$ is the number of degrees of freedom in space. The latter typically includes the number of sites in space (in the context of QCD one should also take into account the number of colors and flavors). The construction (\ref{eq:S_eff}) is the same both for lattice field theories and interacting tight-binding models in condensed matter physics. The key element of our algorithms is the efficient calculation of the derivatives of the fermionic determinant with respect to the bosonic fields, which is essential for the solution of the GF equations. The derivatives of the logarithm of the fermionic determinant can be computed directly using the simple relation
\begin{eqnarray}
\label{eq:det_der}
 \frac{\partial \ln \det M} {\partial \Phi} =  \Tr\left( { M^{-1} \frac {\partial M}{\partial \Phi} }\right).
 \end{eqnarray}
 It turns out that this requires the knowledge of only a few elements of the fermion propagator $M^{-1}$, since the bosonic fields $\Phi$ enter the fermionic operator $M$ locally. 
 
 In the following considerations we rely on the special band structure of the fermionic operator. 
 We start with unimproved staggered fermions, whose fermionic operator can be written as
 \begin{eqnarray}
\label{eq:staggered_action}
 M^{st}_{i,j} = 2 a m \delta_{i,j}  + \nonumber \\ (\eta_{i, 1} e^{\mu a}  U_{i,1} \delta_{i+\hat 1, j} -
 \eta_{j, 1}  U^\dag_{j, 1} e^{-\mu a} \delta_{i-\hat 1, j})  +  \nonumber \\  \sum_{\nu=2}^4 ( \eta_{i, \nu}  U_{i, \nu} \delta_{i+\hat\nu, j} - \eta_{j, \nu}  U^\dag_{j, \nu} \delta_{i-\hat\nu, j})
 \end{eqnarray}
 with the usual staggered phases $\eta_{i, \nu}=(-1)^{i_1+...+i_{\nu-1}}$ and gauge fields $U_{i,\nu}$.
 Here $\mu$ is the chemical potential and $m$ is the mass of fermions, and both are multiplied by the lattice spacing $a$.
 The four-dimensional index $i=(t,x)$ consists of both the temporal $t$ and the three-dimensional spatial part $x$. It is convenient to introduce the spatial part of the fermionic operator $B_t$, which contains all elements of the matrix (\ref{eq:staggered_action}), diagonal in Euclidean time direction for a given time slice $t$. After doing so, (\ref{eq:staggered_action}) can be rewritten as a block matrix consisting of blocks $N_s\times N_s$:
 \begin{widetext}
 \begin{align}
\label{eq:fermionmatrix_staggered}
 &M^{st}(U) 
 = \nonumber \\ &
 \left(
  \begin{array}{ccccccc}
     B_1                            &    e^{\mu a} [U_{(1,x),1}]                 &  0                       & 0      &  0                                        & \ldots                                  &  e^{-\mu a} [U^\dag_{(N_{\tau},x),1}]  \\
     -e^{-\mu a} [U^\dag_{(1,x),1}] &                  B_2                    &  e^{\mu a} [U_{(2,x),1}]     & 0      &  0                                       & \ldots                                  & 0       \\
     0                              &   -e^{-\mu a} [U^\dag_{(2,x),1}]           & B_3                      & \ddots &  0                                        & \ldots                                  & 0       \\
     0                              &  0                                      & \ddots                   & \ddots & \ddots                                    & \ldots                                  & 0       \\
     \vdots                         &                                         &                          & \ddots & \ddots                                    & \ddots                                  & \vdots  \\
     0                              &  0                                      & \ldots                   & 0      & -e^{-\mu a}[U^\dag_{(N_{\tau-2},x),1}]      & B_{N_\tau-1}                            &  e^{\mu a}[U_{(N_{\tau-1},x),1}]   \\
     -e^{\mu a}[U_{(N_{\tau},x),1}]  &  0                                      & \ldots                   & 0      &         0                                 & -e^{-\mu a}[U^\dag_{(N_{\tau-1},x),1}]    & B_{N_\tau}   \\
  \end{array}
 \right).
\end{align}
\end{widetext}
$[U_{(t,x),1}]$ is a diagonal matrix which contains on the main diagonal all gauge field exponents in the Euclidean time direction for a given timeslice $t$.
Following \cite{HASENFRATZ1992539}, the determinant of (\ref{eq:fermionmatrix_staggered}) is equivalent to the determinant of the following matrix:
\begin{align}
\label{eq:fermionmatrix_general}
 &{\overline{M}}^{st}(U)  = \nonumber \\ &
 \left(
  \begin{array}{cccccc}
     1          & D_1 & 0        & 0 & 0 &\ldots  \\
     0          & 1        & D_2 & 0 & 0 & \ldots  \\
     0          & 0        & 1                & D_3 & 0 &    \ldots\\
     0          & 0 & 0     & 1                & D_4 &    \ldots\\
         \vdots & & & & \ddots &                               \\
     -D_{2 N_\tau} & 0  & 0 &       & \ldots   & 1       \\
  \end{array}
 \right).
\end{align}
Now all blocks are of the size $2N_s \times 2N_s$: 
\begin{eqnarray}
D_{2k} &=&  \left(
  \begin{array}{cc}
      e^{\mu a} [U_{(k,x),1}]           &  0 \\
       0                       &  e^{\mu a} [U_{(k,x),1}]      
  \end{array}
 \right), \nonumber \\
D_{2k-1} &=&  \left(
  \begin{array}{cc}
     B_k                     &  I \\
     I                       &  0      
  \end{array}
 \right), \, k=1...N_\tau
 \label{eq:d_blocks_staggered}
 \end{eqnarray}
The same general form of the fermionic operator is also common for interacting tight-binding models in condensed matter physics. See \cite{Conformal_PhysRevB.99.205434} and references therein for more details. The only difference is that the blocks would be of the size $N_s \times N_s$ in the case of an interacting tight-binding model, and their internal construction is also different. However, these details are not important for the present discussion.

The inverse fermionic matrix can also be written in terms of spatial $2N_s \times 2N_s$ blocks,
\begin{align}
\label{eq:propagator_general}
 &{\overline{M}^{st}}^{-1}(U) 
 = \nonumber \\ &
 \left(
  \begin{array}{cccccc}
     g_1          & \ldots & \ldots        & \ldots & \ldots & \bar g_{2 N_\tau}  \\
     \bar g_1          & g_2        & \ldots & \ldots & \ldots & \ldots  \\
     \ldots         & \bar g_2        & g_3                & \ldots & \ldots &    \ldots\\
     \ldots          & \ldots & \bar g_3     & g_4                & \ldots &    \ldots\\
         \vdots & & & & \ddots &                               \\
     \ldots & \ldots  & \ldots &       & \ldots   & g_{2 N_\tau}       \\
  \end{array}
 \right).
\end{align}
The matrix ${\overline{M}^{st}}^{-1}(U)$ is dense, but here we explicitly show only those blocks which are needed for our calculations. In fact, in the trace in (\ref{eq:det_der}), only the off-diagonal blocks ,$\bar g_{n}, n=1...N_\tau$, will contribute to the exact derivatives. Following \cite{Conformal_PhysRevB.99.205434}, an iterative procedure can be used to compute all needed elements of the fermionic propagator,
\begin{eqnarray}
\label{eq:g_off_diag_iter}
\bar g_{i+1}= D_{i+1}^{-1} \bar g_i D_i.
\end{eqnarray}
Once we know one off-diagonal block $\bar g_{n}$ for some $n$, we can, in principle, reconstruct all of them. Of course, we need to invert the $D_i$ blocks after each update of the gauge fields, but taking into account their sparsity (\ref{eq:d_blocks_staggered}), it costs no more than $N_s^3$ operations for each block. In practice, these iterations typically can not last for more than $N_{\rm {Schur}} \sim 10$ time slices due to accumulation of round-off errors. Thus, we compute the Green's functions $\bar g_{n}$ from scratch for each $n=k N_{\rm {Schur}},\ k=0,1,2...$ and use iterations (\ref{eq:g_off_diag_iter}) only in between for intermediate time slices. The Schur complement solver \cite{ULYBYSHEV2019118} is used for finding  $\bar g_{n}$ from scratch, with additional simplifications described in \cite{Conformal_PhysRevB.99.205434}. The solver, including the iterations (\ref{eq:g_off_diag_iter}), scales as $N_s^3 N_\tau$. However, despite the scaling being worse than that of iterative solvers, the method still gives substantial speedup in comparison with the calculation of fermionic determinant using stochastic estimators. The reason is twofold: 1) there is a very small prefactor in the scaling relation, which compensates for the $N_s^3$ term at least for lattices up to $N_s \sim 10^3$; 2) there is no need to repeatedly find solutions for multiple stochastic estimators, since we get all exact derivatives after one application of the Schur solver accompanied with the propagation through the entire Euclidean time extent of the lattice according to (\ref{eq:g_off_diag_iter}). A more careful analysis of the performance of the Schur solver and some benchmarks against an iterative solver were done in \cite{ULYBYSHEV2019118}.
 
The situation is a bit more complicated for Wilson fermions. In this case, we used the derivation of the compressed form of the Wilson fermionic operator from \cite{PhysRevD.82.094027}. Disregarding the constant determinant of the permutation matrix, the determinant of the Wilson fermionic operator is equivalent to the determinant of the following matrix,
\begin{align}
\label{eq:fermionmatrix_wilson}
 &{\overline{M}}^{W}  = \nonumber \\ &
 \left(
  \begin{array}{cccccc}
     A_1                   & C_1           &       0        & 0       &  \ldots       & 0        \\
     0                     & A_2           &  C_2           & 0       &  \ldots       & 0        \\
     0                     & 0             & A_3            & \ddots  &  \ldots       & 0        \\
     \vdots                &               &                & \ddots  &               & \vdots   \\
     0                     & 0             &       \ldots   &         &  A_{N_\tau-1} & C_{N_\tau-1}    \\
     -C_{N_\tau}           & 0             &       \ldots   &         &       0       & A_{N_\tau} \\
  \end{array}
 \right).
\end{align}
$A$ and $C$ are blocks of size $4 N_s \times 4 N_s$. A factor of four appeared due to the Dirac index of Wilson fermions, and $N_s$ includes both the number of sites in space and color degree of freedom as in the case of staggered fermions. We refer to  \cite{PhysRevD.82.094027} for the derivation of this relation and for the exact form of the blocks $A$ and $C$. The gauge fields enter in both the diagonal and off-diagonal blocks, although the diagonal blocks $A_i$ include only spatial links, while the off-diagonal blocks $C_i$ include both spatial links and the links in Euclidean time direction. The determinant of the operator (\ref{eq:fermionmatrix_wilson}) can be simplified further as,
\begin{equation}
\det {\overline{M}}^{W} =  \det Q^W \prod_i \det A_i,
\end{equation}
where
\begin{align}
\label{eq:q_matrix_wilson}
 &{{Q}}^{W}  = \nonumber \\ &
 \left(
  \begin{array}{cccccc}
     I                     & {A_1}^{-1}    &       0        & 0       &  \ldots       & 0        \\
     0                     & I             &  C_1           & 0       &  \ldots       & 0        \\
     0                     & 0             &  I             &{A_2}^{-1}&  \ldots      & 0        \\
     \vdots                &               &                & \ddots  &  \ddots       & \vdots   \\
     0                     & 0             &       \ldots   &         &  I            & A_{N_\tau}^{-1}    \\
     -C_{N_\tau}           & 0             &       \ldots   &         &       0       & I \\
  \end{array}
 \right).
\end{align}
${Q}^{W}$ has exactly the form needed for the Schur solver (\ref{eq:fermionmatrix_general}), thus the blocks of $(Q^W)^{-1}$ essential for the derivatives can be computed with exactly the same algorithm  as described above with only one substitution,
\begin{eqnarray}
D_{2k}=C_k, \, D_{2k-1}={A_k}^{-1}, \, k=1...N_\tau.
\end{eqnarray}
Equation (\ref{eq:det_der}) is also modified. For spatial links $\nu=2,3,4$ it can be written as
\begin{eqnarray}
& \frac{\partial \ln \det  {\overline{M}}^{W}  } {\partial  U_{(t,x),\nu} } =  \Tr\left( { \bar g_{2t} \frac {\partial C_t}{\partial U_{(t,x),\nu}} }\right) -  \\ & \Tr\left(  \bar g_{2t-1} {A_t}^{-1}{ \frac  {\partial A_t}{\partial U_{(t,x),\nu}} {A_t}^{-1} }\right) +  \Tr\left( { {A_t}^{-1} \frac {\partial A_t}{\partial U_{(t,x),\nu}} }\right). \nonumber
\end{eqnarray}
For the links in the Euclidean time direction, the expression is a bit simpler, since they do not enter into the $A_i$ blocks,
\begin{eqnarray}
\frac{\partial \ln \det  {\overline{M}}^{W}  } {\partial  U_{(t,x),1} } =  \Tr\left( { \bar g_{2t} \frac {\partial C_t}{\partial U_{(t,x),1}} }\right).
\end{eqnarray}
In both expressions above, $\bar g_{i}$ are the off-diagonal blocks of $(Q^W)^{-1}$, enumerated according to the convention in (\ref{eq:propagator_general}).
Despite the slightly more complicated expressions for Wilson fermions, one needs only to invert the spatial blocks and apply the Schur complement solver to compute all exact derivatives of the fermionic determinant. 
Thus, the scaling of the method is also $N_s^3 N_\tau$.

\section{\label{sec:model}The model}

\begin{figure}
        \centering
        \includegraphics[scale=0.25, angle=0]{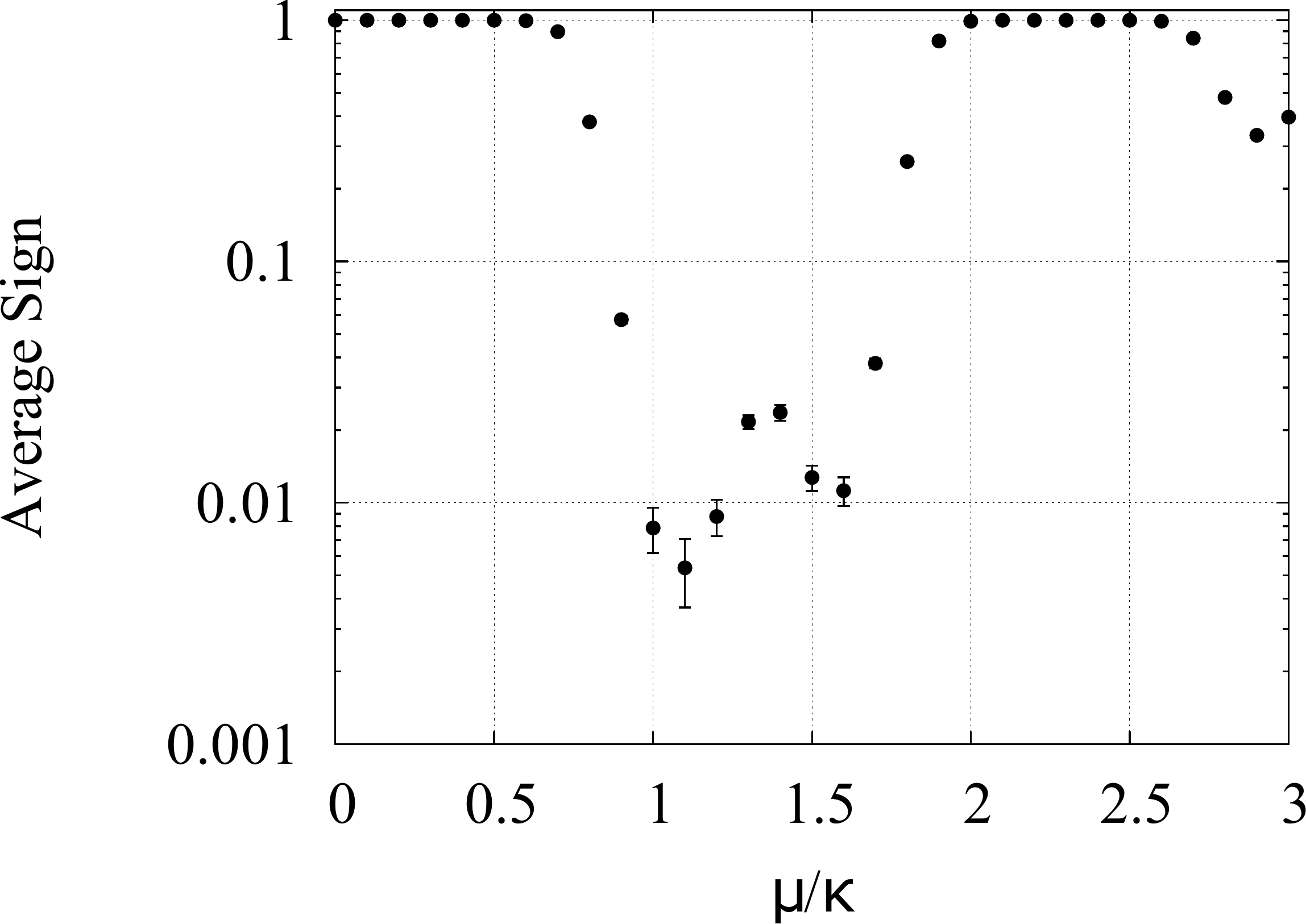}
        \caption{Average sign in BSS-QMC algorithm, taken from the ALF (Algorithm for lattice fermions) package~\cite{Blankenbecler:1981jt,ALF2017}. The calculations were peformed on a hexagonal $4\times4$ lattice with $N_{\tau}=256$ and $\beta=20.0$; $U=2.0 \kappa$. The discrete auxiliary field is coupled to spin as this setup corresponds to the minimal sign problem in BSS-QMC. }
        \label{fig:BSS_QMC_mu_sign}
\end{figure}

In order to demonstrate the performance of the described algorithms, we consider the Hubbard model on the hexagonal lattice at finite chemical potential. At half-filling, this model is known to exhibit a semimetal-to-insulator transition (\cite{AssaadHerbut2013,Sorella2012}), which gives us an appropriate scale for the interaction strength. Furthermore, the particle-hole symmetry at half-filling helps to identify and characterize the thimbles and saddle points before we increase the chemical potential. We start from the form of the Hamiltonian written in the particle-hole basis in order to have a manifestly positive-definite weight for the Hubbard field configurations at half filling
\begin{eqnarray}
  \label{Hamiltonian_el_hol}
  \!\!\!\hat{{H}}\! =\! -\kappa \sum_{\langle x,y\rangle} (  \hat a^\dag_{x} \hat a_{y} \!+\! \hat b^\dag_{x} \hat b_{y} \!+\! \mbox{h.c} ) \!+\! \frac{U}{2} \sum_{x} \hat q_x^2 \!+\! \mu  \sum_x \hat q_x,
\end{eqnarray}
where $\hat a^\dag_{x}$ and  $\hat b^\dag_{x}$ are creation operators for electrons and holes,  $\hat q_x=\hat n_{x, \text{el.}} - \hat n_{x, \text{h.}}={\hat a^\dag_{x}} \hat a_{x}- {\hat b^\dag_{x}} \hat b_{x}$ is the charge operator, $\kappa$ is the hopping parameter, $U>0$ is the Hubbard interaction, and $\mu$ is the chemical potential. Due to the van Hove singularity present in the density of states, one can clearly identify the scale where new physics is expected: $\mu=\kappa$ (see \cite{PhysRevLett.104.136803, Korner:2017qhf} and references therein). Special attention will be paid to this value of $\mu$ in the calculations that follow, since it is also the region with the most severe sign problem \cite{PhysRevB.97.075127}, as can be seen from the test calculations made with BSS-QMC (see Fig.~\ref{fig:BSS_QMC_mu_sign}).  
Interaction strength is chosen in order to probe both sides of the antiferromagnetic (AFM) phase transition, which happens at $U \approx 3.8\kappa$ \cite{AssaadHerbut2013,Buividovich:2018yar} on hexagonal lattice. Thus, we usually take tree values of $U$: $U=2\kappa$ of $U=3\kappa$ in semi-metallic phase, $U=3.8 \kappa$ at the phase transition and $U=5\kappa$ in AFM phase. Temperature is chosen to be below the Neel temperature for $U=5\kappa$, so that we indeed are in ordered phase at that value of interaction strength \cite{Buividovich:2018yar}. Also, in this region of temperatures ($\beta \kappa=20...30$), the BSS-QMC method experiences the exponential decay of the average sign, reaching as small values as $10^{-2}...10^{-3}$ even on small $2\times2$ lattices, as it is shown in Fig.~\ref{fig:BSS_QMC_mu_sign}. The last argument, why we think this temperature is small enough stems from comparison with other recent papers dealing with QMC calculations for doped Hubbard model: $\beta \kappa=5$ for BSS-QMC calculations in \cite{Huang987} and $\beta \kappa=3$ in tempered Lefschetz thimbles method \cite{Fukuma:2019wbv}. This comparison shows that in our saddle point analysis we could reach temperatures well below those achieved in the state-of-the-art QMC computations. 

The next step in constructing the path integral formulation of the model is to introduce the Trotter decomposition:
\begin{equation}
e^{-\beta \hat{{H}}} \approx ... e^{-\delta \hat{{K}}} e^{-\delta \hat{{H}}_U} e^{-\delta \hat{{K}}}e^{-\delta \hat{{H}}_U} ... + O(\delta^2)
\end{equation}
where $\hat{{K}}$ is the collection of all bilinear fermionic terms in $\hat{{H}}$, and $\hat{{H}}_U$ is the interaction part of the full Hamiltonian. Here we have introduced $\delta$, which specifies the discretization of Euclidean time, $N_\tau \delta=\beta$, where $N_\tau$ constitutes the Euclidean time extent of the lattice. Below, we will refer to $\beta$ in the units of inverse hopping. 

One can obtain an additional, nonphysical, degree of freedom in the Hamiltonian, by applying the following identity to the interaction term
\begin{eqnarray}
\frac{U}{2}\hat q_x^2 = \frac {\alpha U}{2}\hat q_x^2 - \frac{(1-\alpha) U}{2} \hat s_x^2 + (1-\alpha) U \hat s_x,
\label{eq:split_int}
\end{eqnarray}
where $\hat s_x = \hat n_{x, \text{el.}} + \hat n_{x, \text{h.}}$ is the spin operator. We can now simultaneously introduce two continuous auxiliary fields by applying the standard Hubbard-Stratonovich transformations to each four-fermion term in (\ref{eq:split_int}),
\begin{align}
\label{continuous_HS_imag}
  e^{-\frac{\delta}{2}\sum\limits_{x,y} U_{x,y} \hat n_x \hat n_y} \! \cong \! \int \! D \phi_x\, e^{- \frac{1}{2\delta} \sum\limits_{x,y} \phi_x U^{-1}_{xy} \phi_y} e^{i \sum\limits_x \phi_x \hat n_x}, \\
   e^{\frac{\delta}{2}\sum\limits_{x,y} U_{x,y} \hat n_x \hat n_y} \! \cong \! \int \! D \phi_x\, e^{- \frac{1}{2\delta} \sum\limits_{x,y} \phi_x U^{-1}_{xy} \phi_y} e^{ \sum\limits_x \phi_x \hat n_x}.
\label{continuous_HS_real}
\end{align}
The first four-fermionic term can be transformed into a bilinear using (\ref{continuous_HS_imag}), and the second using (\ref{continuous_HS_real}). This is not the most general possible decomposition of four-fermionic terms into bilinear ones, but the one most commonly used in QMC algorithms with continuous auxiliary fields. This representation was first proposed in \cite{complex1} and was also used in several recent papers \cite{Assaad_complex, Ulybyshev:2017}.
The parameter $\alpha \in [0,1]$ defines the balance between auxiliary fields coupled to the charge ($\hat q_x$)  and spin ($\hat s_x$) density. This particular representation has an important advantage over discrete auxiliary fields in that it also works for non-local interactions, so that we do not need to introduce a new auxiliary field for every pair of interacting electrons. 

The details of the construction of the path integral are straightforward and can be found in \cite{Ulybyshev:2013swa,SmithVonSmekal,Assaad_complex}. Here we simply state the explicit form of the partition function which we have used in our calculations:
\begin{align}
  &\mathcal{Z}\!=\! \int\!\mathcal{D} \phi_{x,\tau}\, \mathcal{D} \chi_{x,\tau}\, e^ {-S_{\alpha}}  \det M_{\text{el.}} \det M_{\text{h.}},&  \\
   &S_\alpha[\phi_{x,\tau},\chi_{x,\tau}] \!=\! \sum_{x,\tau}  \left[\frac  {\phi_{x,\tau}^2} {2 \alpha \delta U}  \!+\!  \frac {(\chi_{x,\tau}\!-\! (1\!-\!\alpha) \delta U)^2} {2 (1\!-\!\alpha) \delta U}\right],&\nonumber
  \label{eq:Z_continuous}
\end{align}
where the fermionic operators are given by
\begin{eqnarray}
 M_{\text{el.,h.}} = I +\prod^{N_{\tau}}_{\tau=1} \left[{ e^{-\delta \left(h\pm\mu\right)} \diag{ e^{\pm i \phi_{x,\tau}+\chi_{x,\tau}} } }\right]. 
  \label{eq:M_continuous}
\end{eqnarray}
This fermionic operator can be rewritten in the form (\ref{eq:fermionmatrix_general}), where 
\begin{eqnarray}
D_{2k}=\diag{ e^{\pm i \phi_{x,\tau}+\chi_{x,\tau}} } \\
D_{2k-1}= e^{-\delta \left(h\pm\mu\right)}, k=1...N_\tau. \nonumber
\label{eq:tb_model_substitution}
\end{eqnarray}
Thus, all algorithms for the fast solution of the GF equations, described in the section \ref{sec:algoriths} are fully applicable. 
We denote the field coupled to charge density as $\phi_{x,\tau}$, and the field coupled to spin density as $\chi_{x,\tau}$. The full action, which is used in Monte Carlo sampling, involves both the bosonic action of the auxiliary fields as well as the logarithm of the fermionic determinants,  $S=S_{\alpha} - \ln (\det M_{\text{el.}} \det M_{\text{h.}})$.
The total number of auxiliary fields is equal to $N=2 N_s N_\tau$ if $\alpha \in (0,1)$, so that both fields participate, and $N= N_s N_\tau$ if $\alpha = 0,1$, where only one type of field remains.

\begin{figure}
        \centering
        \includegraphics[scale=0.3, angle=0]{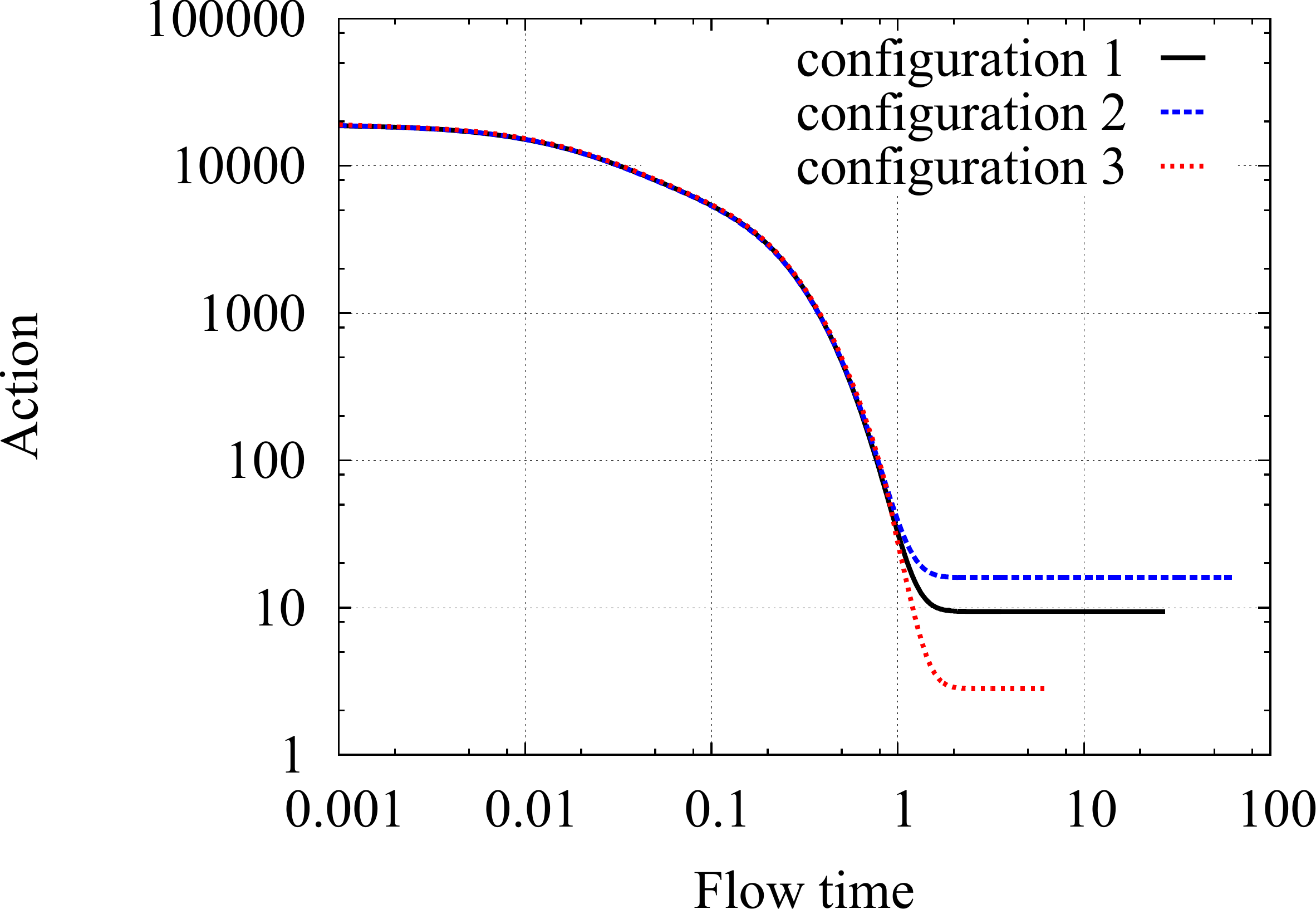}
        \caption{An illustration of the downward gradient flow procedure for three thermalized configurations belonging to different thimbles at half filling. The plot shows the evolution of the action with the flow time. The ensemble consists of a $6\times6$ lattice with $N_{\tau}=256$ and $\beta=20.0$, $U=5.0 \kappa$, $\alpha=0.9$. One can clearly see how the configurations end up at three different saddle points after completion of the flow.}
        \label{fig:real_action_flow}
\end{figure}

\section{\label{sec:spstudy}Saddle points study}
\subsection{\label{subsec:spstudy:halffil}Saddle points at half-filling}

Our goal is to study realistic lattice volumes in order to get a quantitative idea of what the thimbles decomposition (\ref{eq:thimbles_sum_and_integral}) looks like as we approach both the thermodynamic limit in spatial volume and the continuum limit in Euclidean time. Unfortunately, at large lattice volumes, the fully analytical approach for finding saddle points (as was done in~\cite{Ulybyshev:2017} on lattices with up to four sites) does not work. Thus, in this study we are using a completely different approach which is based on importance sampling and fast solutions of the GF equations, using the calculations of the derivatives of the fermionic determinant described in section \ref{sec:algoriths}. 

At half-filling, this method starts with the generation of lattice configurations using standard hybrid Monte Carlo (HMC) techniques.  After this, we numerically integrate the GF equations for each field configuration for a finite flow time, in order to reach the local minimum of the action. At half filling, when thimbles are bounded within $\mathbb{R}^N$, the local minimum of the action always corresponds to a relevant saddle point. At the end of this sequence of steps, the distribution of lattice ensembles, taken after employing the GF procedure, gives an accurate characterization of the relevant saddle points at half-filling if the initial set of configurations was ergodic. An example of such a process is shown in Fig.~\ref{fig:real_action_flow}. After generating configurations using HMC, one can observe the approach to the saddle point in our gradient flow routine. As noted, the real part of the action should monotonically decrease and eventually, at a certain flow time, converge to the value at the saddle. In general, the method scales similar to the Schur complement solver as $N_s^3 N_\tau$.

\begin{figure}
        \centering
        \includegraphics[width=3.5in, angle=0]{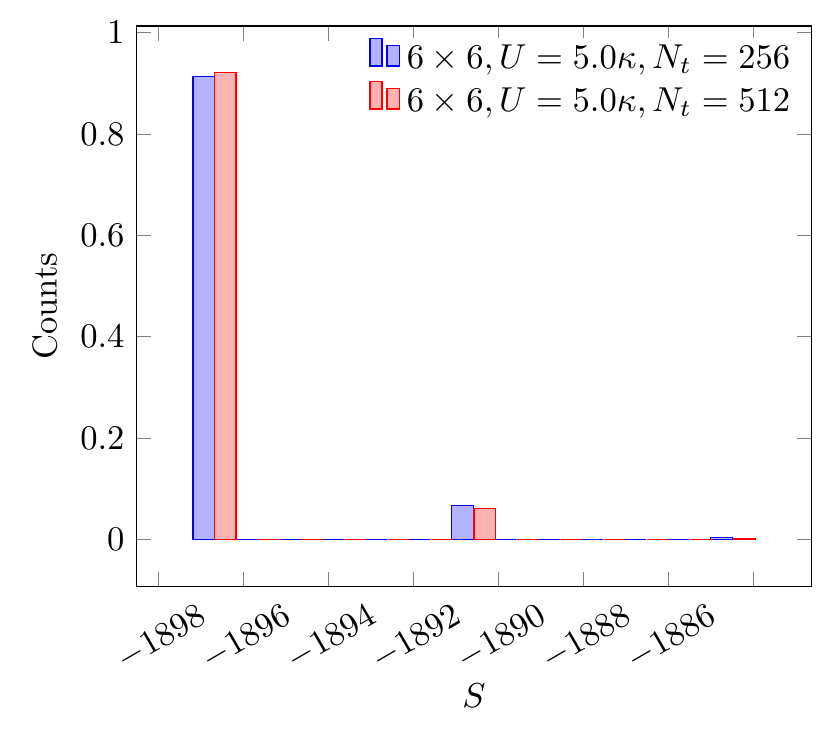}
        \caption{Distribution of the saddle points at half filling for a $6\times6$ lattice in the strong coupling regime with $U=5.0 \kappa$. Two cases are compared: (LHS) $N_{\tau}=256$ and (RHS) $N_{\tau}=512$ for fixed temperature $\beta=20.0$ and $\alpha=0.9$. One can see that the distribution is almost identical and thus we can claim that we are close close enough to the continuum limit in the Euclidean time direction.}
        \label{fig:two_dt_histogram}
\end{figure}

\begin{figure}
        \centering
        \includegraphics[width=3.5in, angle=0]{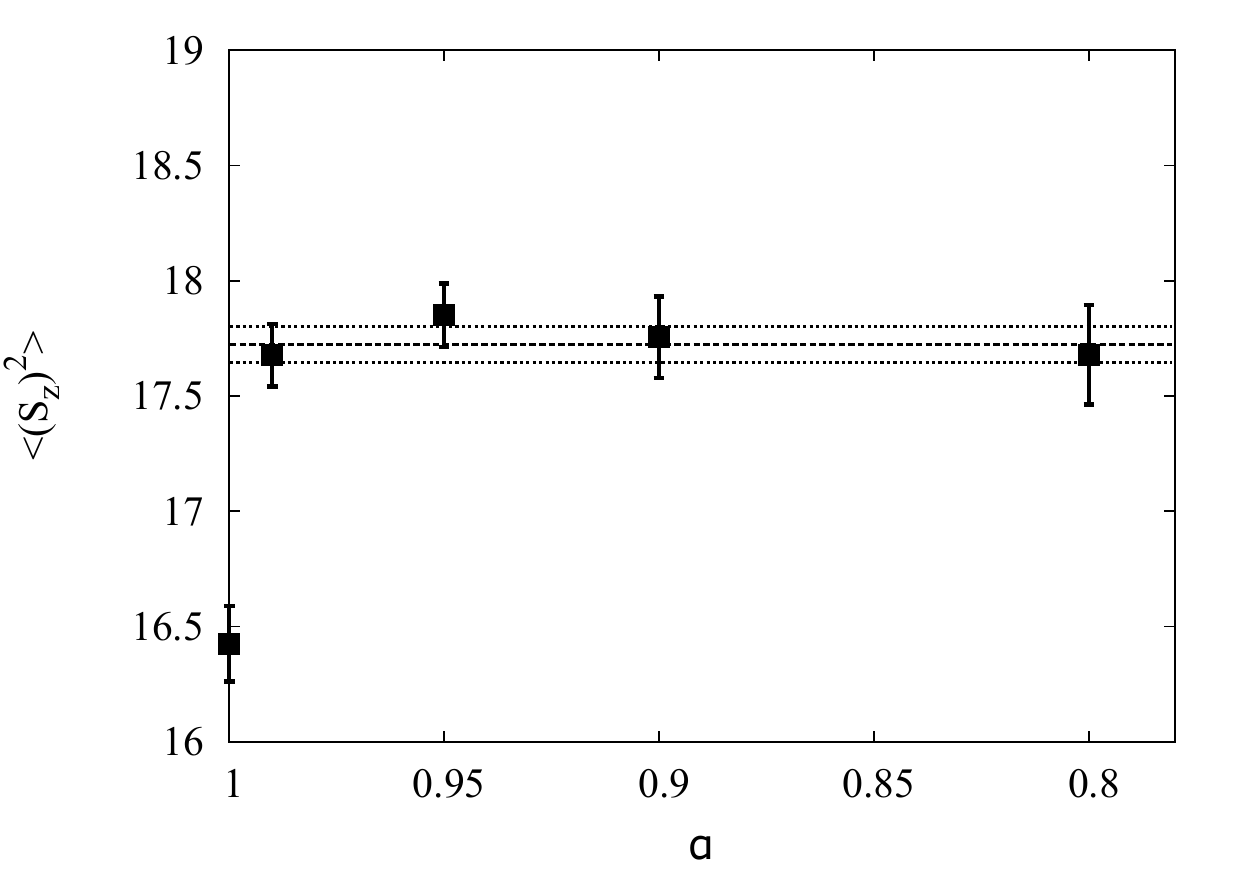}
        \caption{The dependence of the squared spin at one sublattice (see eq.~(\ref{eq:spin_squared})) on $\alpha$. The observable is computed on $6\times6$ lattice with $N_{\tau}=128$ and $\beta=20.0$, $U=3.8 \kappa$. The value from BSS-QMC is shown with the dashed line which is representing the mean value and the dotted lines are representing the errorbars.}
        \label{fig:Sz_alpha}
\end{figure}

A possible source of systematic error in our lattice set up is the discretization in Euclidean time that results from the Trotter decomposition. Thus, we first checked that we have already effectively arrived at the continuum limit in Euclidean time. In Fig.~\ref{fig:two_dt_histogram}, the plot shows the histogram of the distribution of the action for the field configurations after GF. As the initial configurations were generated using HMC, the height of each bar corresponds to the exact weight of the thimble attached to the corresponding saddle point whose value of the action is denoted by the position of the bar. 
In Fig.~\ref{fig:two_dt_histogram} we display the histograms for two lattice spacings at fixed $\beta$. The results are almost identical, and thus we can claim that with $N_\tau=256$ at $\beta=20$, we are already close enough to the continuum limit in Euclidean time. This gives us confidence that our study of the features of the saddle points and thimbles is independent of the step size in Euclidean time. We will use the same style of plots to characterize the structure of the thimbles decomposition below.

\begin{figure}
        \centering
        \includegraphics[width=3.5in, angle=0]{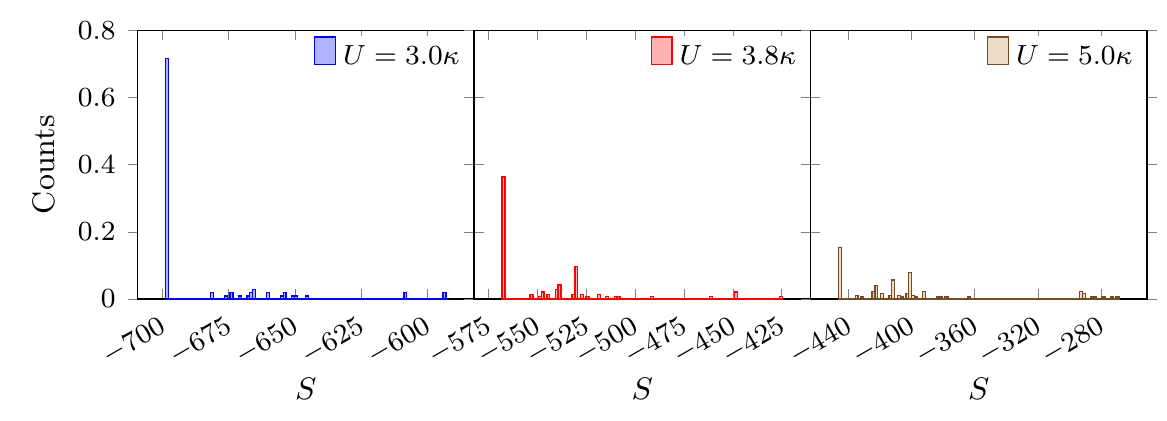}
        \caption{The distribution of the action of saddle point configurations at half filling for $\alpha=0.01$. The ensembles consist of a $6\times6$ lattice with $N_{\tau}=256$ and $\beta=20.0$, and three different values of interaction strength: (left) $U=3.0 \kappa$, (middle) $U=3.8 \kappa$, (right) $U=5.0 \kappa$.}
        \label{fig:small_alpha_histogram}
\end{figure}

We now proceed to study saddle points at different $\alpha$. One important thing to note is that at half-filling, we cannot faithfully sample the path integral at the extreme values $\alpha=1.0$ and $\alpha=0.0$. In both cases (see~\cite{Assaad_complex, Ulybyshev:2017, Wynen:2018ryx}), the product of fermionic determinants is equal to the square of some real-valued function 
\begin{equation}
\label{eq:non_ergodicity}    
\det M_{\text{el.}} \det M_{\text{h.}}|_{\alpha=0,1;\mu=0} = F^2.
\end{equation}
Thus only one constraint, $F=0$, needs to be satisfied in order to have both fermionic determinants equal to zero. It follows that the dimensionality of the manifolds on which the determinant vanishes is equal to $N-1$ and therefore they cut $\mathbb{R}^N$ into disconnected regions. As a result, HMC can not penetrate through these domain walls~\cite{Assaad_complex, Ulybyshev:2017, Wynen:2018ryx}, and we cannot rely on it to generate an ergodic set of configurations. However, as was shown in Fig.~\ref{fig:Sz_alpha} (and one more example will also be shown below), even a small shift of  $\alpha$ from these extreme values is enough to restore ergodicity.
We explicitly check the value of squared spin per sublattice
\begin{equation}
  \langle (S_z)^2 \rangle  = \left\langle \left({ \sum_{i \in \textrm{1st. sublat.}}  \hat S^i_z  } \right)^2  \right\rangle.
  \label{eq:spin_squared}
\end{equation}
 and compare the results from HMC with BSS-QMC, which does not have ergodicity issues due to the formulation in terms of discrete fields. We can thus safely use, e.g. $\alpha=0.01$ and $\alpha=0.99$ in order to gain an understanding of the thimbles decomposition when we have either a dominant spin-coupled field auxiliary field or a dominant charge-coupled auxiliary field, respectively.

\begin{figure}
        \centering
        \includegraphics[scale=0.30, angle=0]{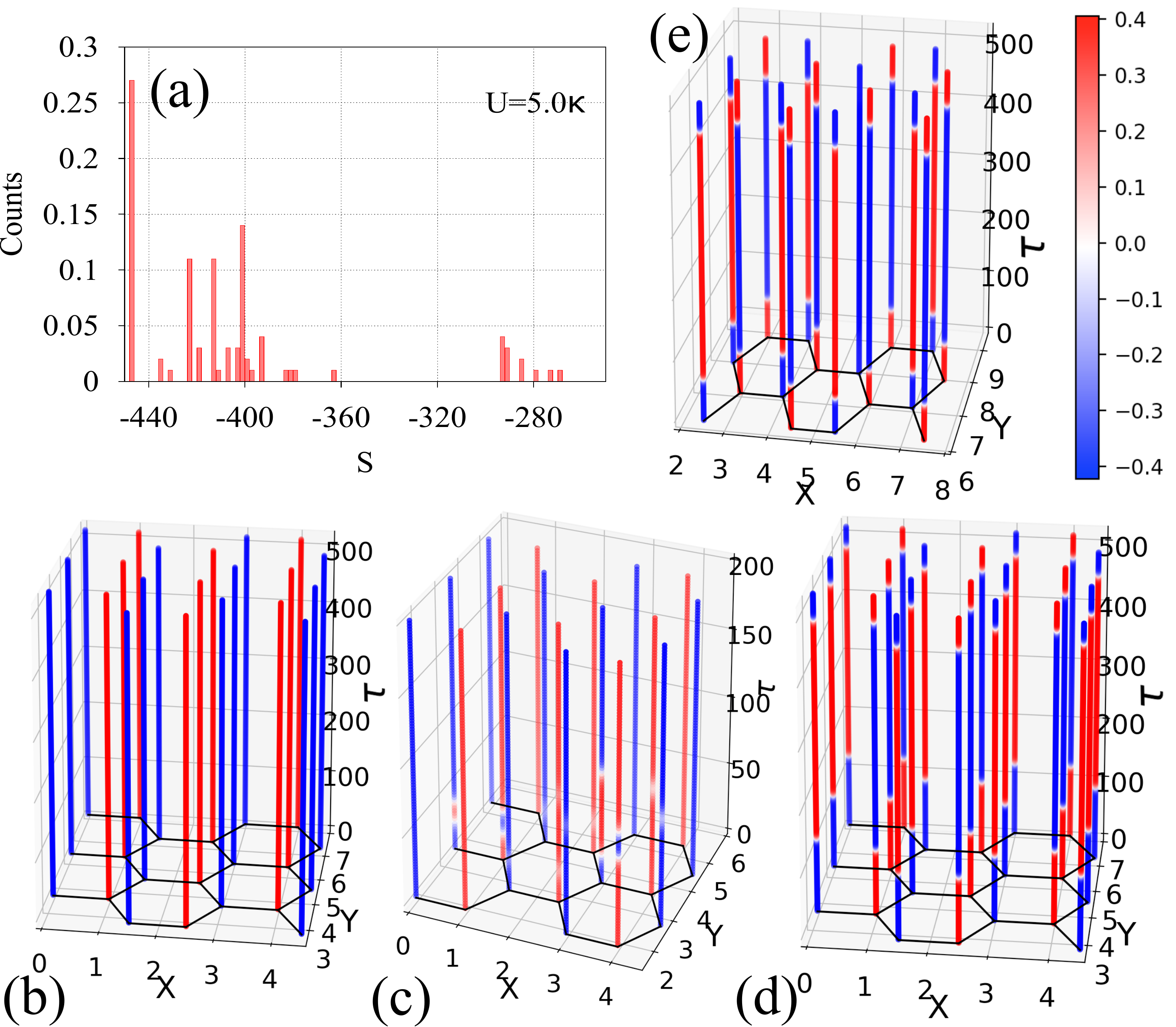}
        \caption{Representative field configurations at saddle points for mostly spin-coupled auxiliary field at half filling ($\alpha=0.01$). The $\chi$-field is shown, while the $\phi$-field is always equal to zero at these saddle points. (a) distribution of action of saddle point configurations for a $6\times6$ lattice with $N_{\tau}=256$ at $U=5.0 \kappa$ and $\beta =20.0$  (we repeat the histogram for $U=5.0 \kappa$  for reference). (b) The AFM mean-field saddle, which corresponds to the bar with the lowest value of the action in the histogram. (c) Local instantons on the background of mean-field vacuum (peaks between $S=-440$ and $S=-400$ in the histogram). (d) Global mean-field instanton (peak at $S \approx -295$ in the histogram). (e) Saddle points with local violations of the structure of mean-field instanton ( $S\approx -280$ in the histogram).
        In all cases, the value of the $\chi$-field is represented by the color of the world lines drawn in the Euclidean time direction emanating from each site on the hexagonal lattice. In (b)-(e), we have only depicted the part of the lattice where interesting features involving the $\chi$-field are found for simplicity. }
        \label{fig:spin_saddles}
\end{figure}

We first study saddle points at $\alpha=0.01$, when the spin-coupled field $\chi_{x,\tau}$ is the most important and $\phi_{x,\tau}$ is always equal to zero at the saddle points. The results are shown in Fig.~\ref{fig:small_alpha_histogram} at different values of the interaction strength, which correspond to the semi-metal (SM) phase ($U=3 \kappa$), the region close to the phase transition ($U=3.8 \kappa$), and the antiferromagnetic (AFM) phase ($U=5.0 \kappa$). In all cases, there is a dominant saddle point corresponding to the smallest value of the action, but its dominance becomes less and and less pronounced as we move towards the AFM phase. A more detailed study of saddles is presented in Fig.~\ref{fig:spin_saddles} for the case of large interaction strength, $U=5 \kappa$. The lowest saddle (Fig.~\ref{fig:spin_saddles}\textcolor{red}{(b)}) is just a static solution which corresponds to $\chi_{x,\tau}=\pm \chi_0$, with the sign depending on the sublattice.
Both these saddles are just two identical mean field solutions corresponding to antiferromagnetic ordering (N\'eel state). They appear as a consequence of the bipartite nature of the lattice. Since there are two stationary vacua, ``instanton" solutions, which represent tunneling events between the two vacua, inevitably appear. We indeed observe these instantons which correspond to saddle points with larger actions, examples of which are shown in Fig.~\ref{fig:spin_saddles}\textcolor{red}{c-e}. In (c), one sees how, at a given site, one can have virtually instantaneous tunneling of the value of $\chi$ between $+\chi_0$ and $-\chi_0$ and back again, where $\chi_0$ is the same value that appears in the mean-field configuration. In these configurations, the tunneling occurs locally both in space and in time. However, there are cases such as (d), where the tunneling from $\pm \chi_0$ to $\mp \chi_0$ occurs all across the lattice in space at some Euclidean time $\tau_0$, and at a later time $\tau_1$, the fields return to their original configuration. Thus, we have two ``global instantons''. Finally, we have observed cases such as (e), where a similar pair of ``global instantons'' exists, with the caveat that the tunneling structure is violated locally in space. The identification of these examples with the action depicted in the histogram is described in the caption to Fig.~\ref{fig:spin_saddles}.

\begin{figure}
        \centering
        \includegraphics[width=3.5in, angle=0]{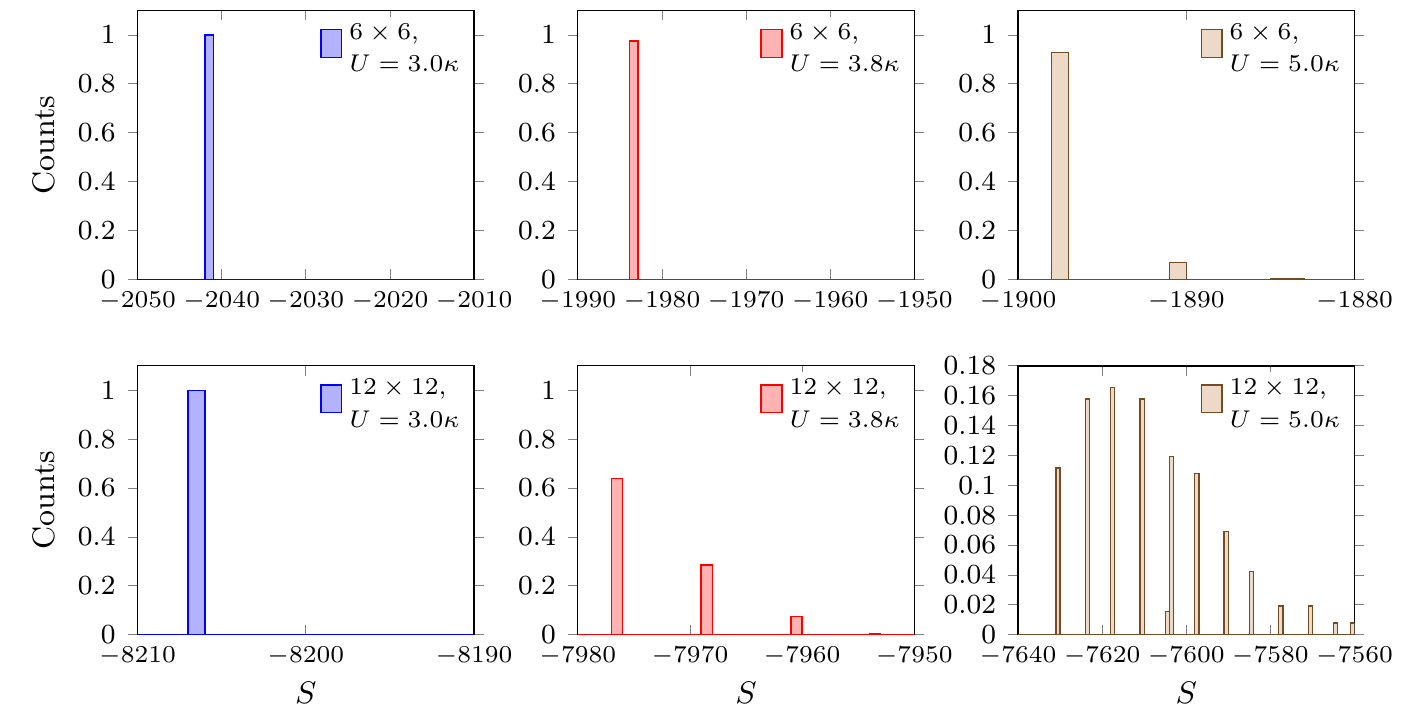}
        \caption{The distribution of the action of saddle point configurations at half filling for an intermediate case, $\alpha=0.9$. The ensembles consist of the following: (upper panel) $6\times6$ and (lower panel) $12\times12$ lattice with $N_{\tau}=256$ and $\beta=20.0$, and three different values of interaction strength: (upper and lower left) $U=3.0 \kappa$; (upper and lower middle) $U=3.8 \kappa$; (upper and lower right) $U=5.0 \kappa$. The histograms reveal a much more regular (in comparison with Fig.~\ref{fig:small_alpha_histogram}) system of saddle points. The lowest saddle points correspond to the vacuum configuration (all auxiliary fields are equal to zero). }
        \label{fig:intermediate_alpha_histogram}
\end{figure}

\begin{figure*}[t]
 \includegraphics[scale=0.55, angle=0]{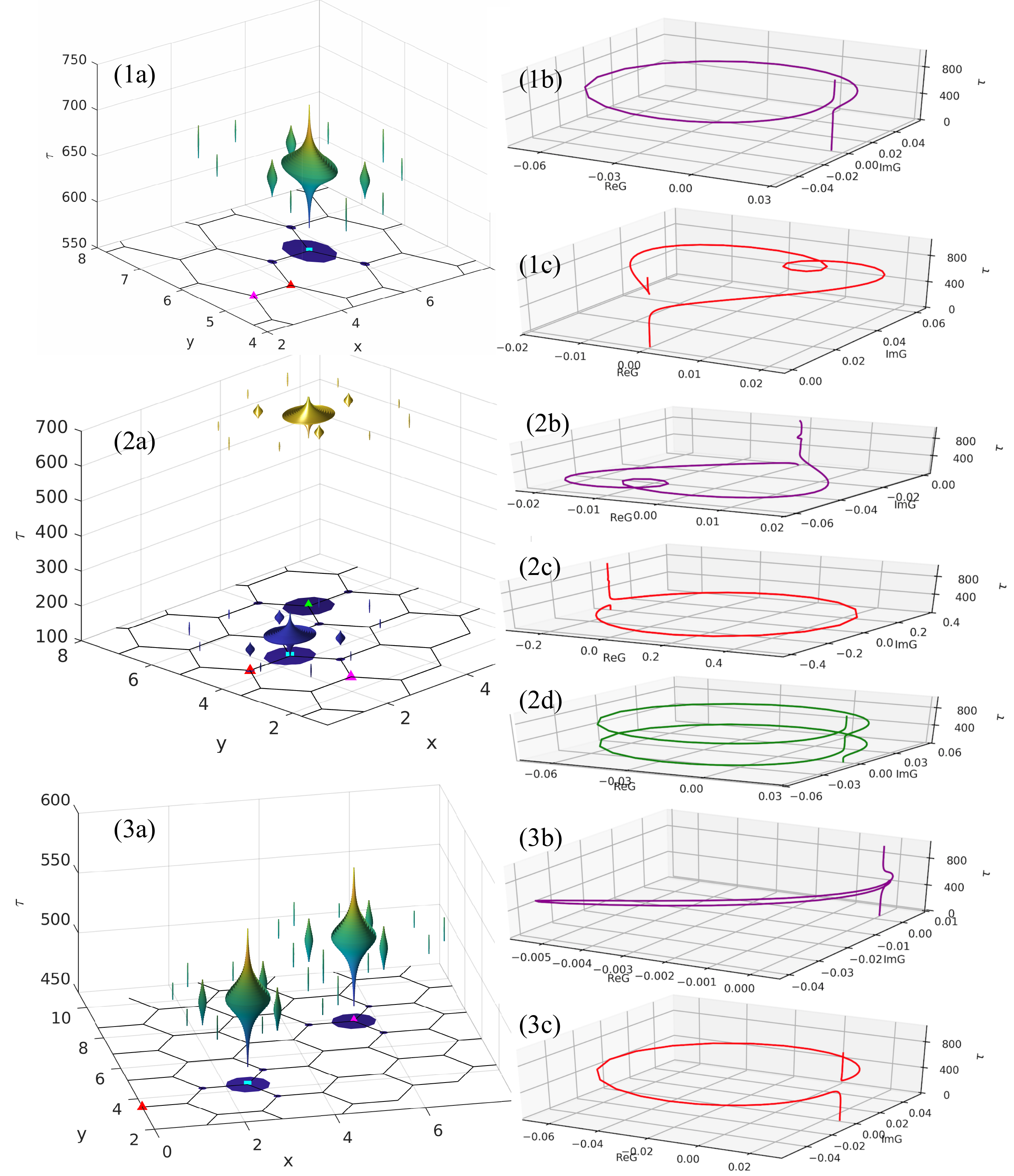}
        \caption{Representative field configurations at saddle points for mostly charge-coupled auxiliary field at half filling($\alpha=0.9$, $6\times6$ lattice with $N_{\tau}=512$ at $U=5.0 \kappa$ and $\beta =20.0$, corresponds to the red histogram in Fig.~\ref{fig:two_dt_histogram}). The $\chi$-field is always equal to zero, while the modulus of the $\phi$-field is shown as the width of a blob at a given spatial lattice site and time step in Euclidean time. For clarity, we only draw world lines if $|\phi|>\epsilon$, with $\epsilon$ some suitably small threshold. In order to make the position of the world lines clear with respect to the spatial lattice, we also draw their projections on the $\tau=0$ plane. The vacuum field configuration corresponds to all fields equal to zero. This saddle corresponds to the bar at lowest action in the red histogram of Fig.~\ref{fig:two_dt_histogram}. (1a)  The lowest non-trivial saddle point corresponds to the bar at $S\approx-1891$ in the histogram \ref{fig:two_dt_histogram}\textcolor{red}{b}. This field configuration is clearly localized, and serves as an elementary quantum to construct further saddle points with higher actions. (2a,3a) Two saddle points which correspond to the third bar in the red histogram of Fig.~\ref{fig:two_dt_histogram} (located at $S\approx-1884$, the bar can not be seen due to the scale). Plots (1b, 1c) show the evolution with $\tau$ of the equal-time fermionic propagator $g(x, y, \tau)$ for the one-blob saddle point shown in (1a). One of the endpoints $x$ is located at the center of the blob (marked with blue square in the projection onto $\tau=0$ plane). The two other endpoints $y$ are marked with a violet and a red triangle in the projection. They correspond to the plots (1b) and (1c) (drawn in the same colors as the corresponding triangles). The same rule is applied to the plots (2b,2c,2d) and (3b,3c): they demonstrate the properties of equal-time fermionic propagators with respect to the saddle points shown in (2a) and the blue histogram of Fig.~\ref{fig:two_dt_histogram}, respectively.}
        \label{fig:charge}
\end{figure*}

 Next, in Fig.~\ref{fig:intermediate_alpha_histogram} we display the results at a larger value of $\alpha$ where the auxiliary field which couples to charge-density starts to dominate and all saddle points are located at $\chi_{x,\tau}=0$. The latter fact automatically ensures that the structure of the saddle points is completely different from the one at small $\alpha$. The histograms for the same three couplings show that the situation improves and that the construction of the saddle points is now more regular since they are equally spaced in action. A comparison of the $6 \times 6$ and $12\times12$ lattices (Fig.~\ref{fig:intermediate_alpha_histogram}\textcolor{red}{(upper panel)} and Fig.~\ref{fig:intermediate_alpha_histogram}\textcolor{red}{(lower panel)}) shows that the number of saddles appearing in the histogram increases with the increasing volume, particularly at larger values of the coupling $U$. However, the general structure of saddle points remains essentially the same. This situation is demonstrated in Fig.~\ref{fig:charge}. Here we have taken a $6\times6$ lattice with $N_\tau=512$ at $U=5 \kappa$, as an example (corresponding histogram is shown in the Fig.~\ref{fig:two_dt_histogram}). However, the same field configurations were observed at saddle points for other $U$, $N_\tau$ and also at a volume of $12\times 12$. For all histograms, shown here for $\alpha=0.9$ (Fig.~\ref{fig:intermediate_alpha_histogram}), the first bar corresponds to the vacuum saddle $\phi_{x,\tau}=\chi_{x,\tau}=0$. The next bar corresponds to the localized field configurations shown in Fig.~\ref{fig:charge}\textcolor{red}{-1(a)}. These localized features come in two types, differing only in the sign of the $\phi_{x,\tau}$ field: $\phi_{x,\tau} \rightarrow -\phi_{x,\tau}$. We will refer to these structures as ``blob'' and ``anti-blob'' in the subsequent discussion. The third bar in the histograms for $\alpha=0.9$ corresponds to the three combinations one can construct out of two of these localized objects: blob-blob, blob-anti-blob and two anti-blobs, where the objects are located at some spatial separation on the lattice. Two examples are shown in Fig.~\ref{fig:charge}\textcolor{red}{-2(a)} and \ref{fig:charge}\textcolor{red}{-3(a)}.
 All further saddle points consist of more complicated combinations of increasing number of blobs/anti-blobs that are localized somewhere within the lattice. 
 The single blob shown in Fig.~\ref{fig:charge}\textcolor{red}{-1(a)} has an action given by $S_1 = S_0 + \Delta S$, where $S_0$ is the action of the trivial vacuum.  Both configurations in Fig.~\ref{fig:charge}\textcolor{red}{-2(a)} and \textcolor{red}{3} have actions given by $S_2 \approx S_0 + 2\Delta S$ to a very high precision.  It follows that the actions of $n$-blob configurations should be concentrated around  $S_n = S_0 + n\Delta S$, with the width of the distribution slightly widening with increasing $n$. This is due to that fact that as the density of blobs increases, they are no longer well-separated and start to interact with each other.

\begin{figure}
        \centering
        \includegraphics[scale=0.3, angle=0]{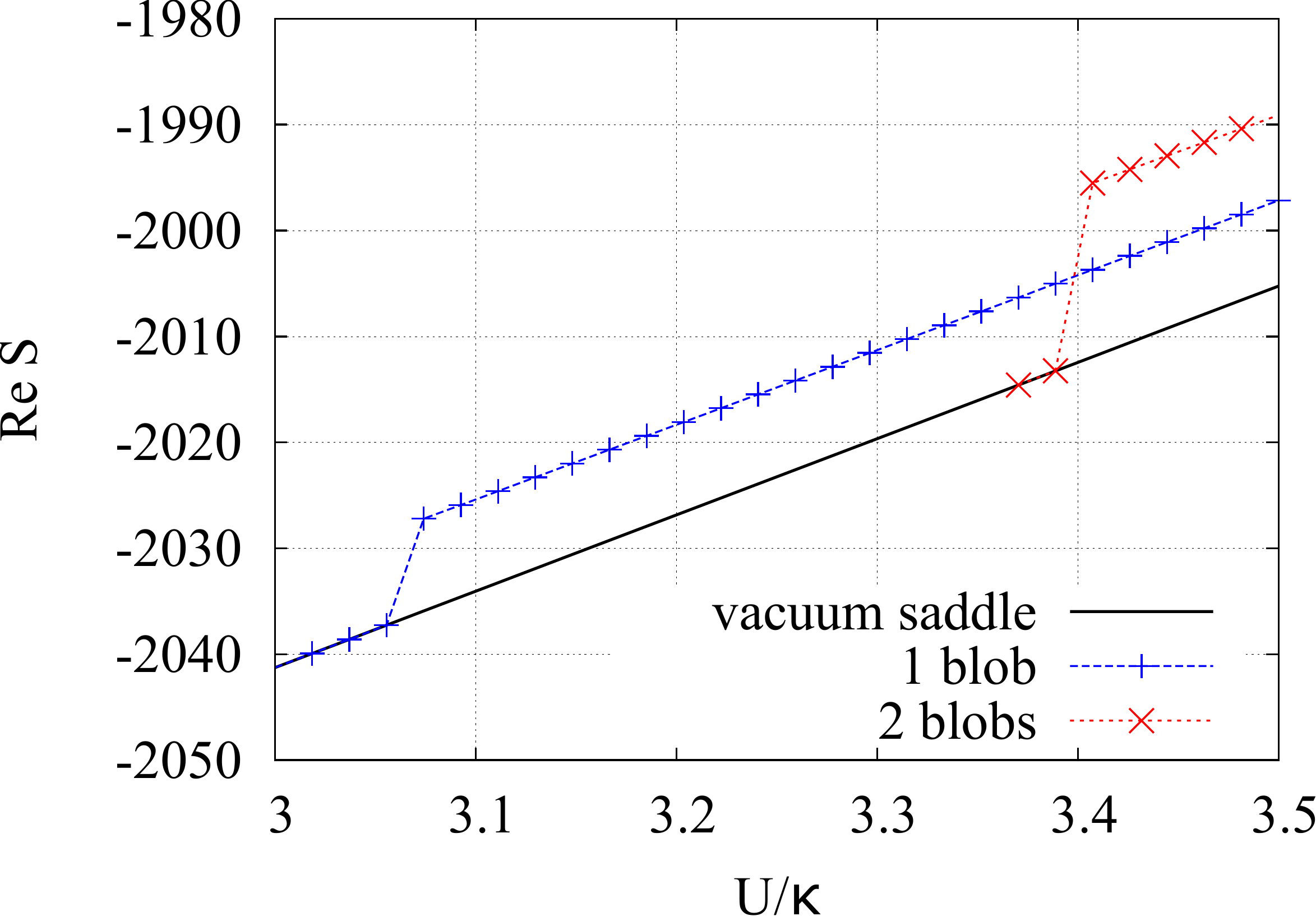}
        \caption{The dependence of the saddle points for mostly charge-coupled auxiliary field on the interaction strength at half filling ($\alpha=0.9$, $6\times6$ lattice with $N_{\tau}=256$ at  $\beta =20.0$). Each subsequent point is obtained via GF from the previous one (moving from larger $U$). If the saddle point becomes irrelevant, the flow shows decays into the vacuum saddle. Due to the localized structure of the field configurations at saddle points, they remain equidistant in action. However, at small interaction strength non trivial saddles decay into the vacuum one. This illustrates the influence of non-trivial saddle points on the physics in the strongly-coupled regime.}
        \label{fig:history_U}
\end{figure}

\begin{figure}
        \centering
        \includegraphics[scale=0.2, angle=270]{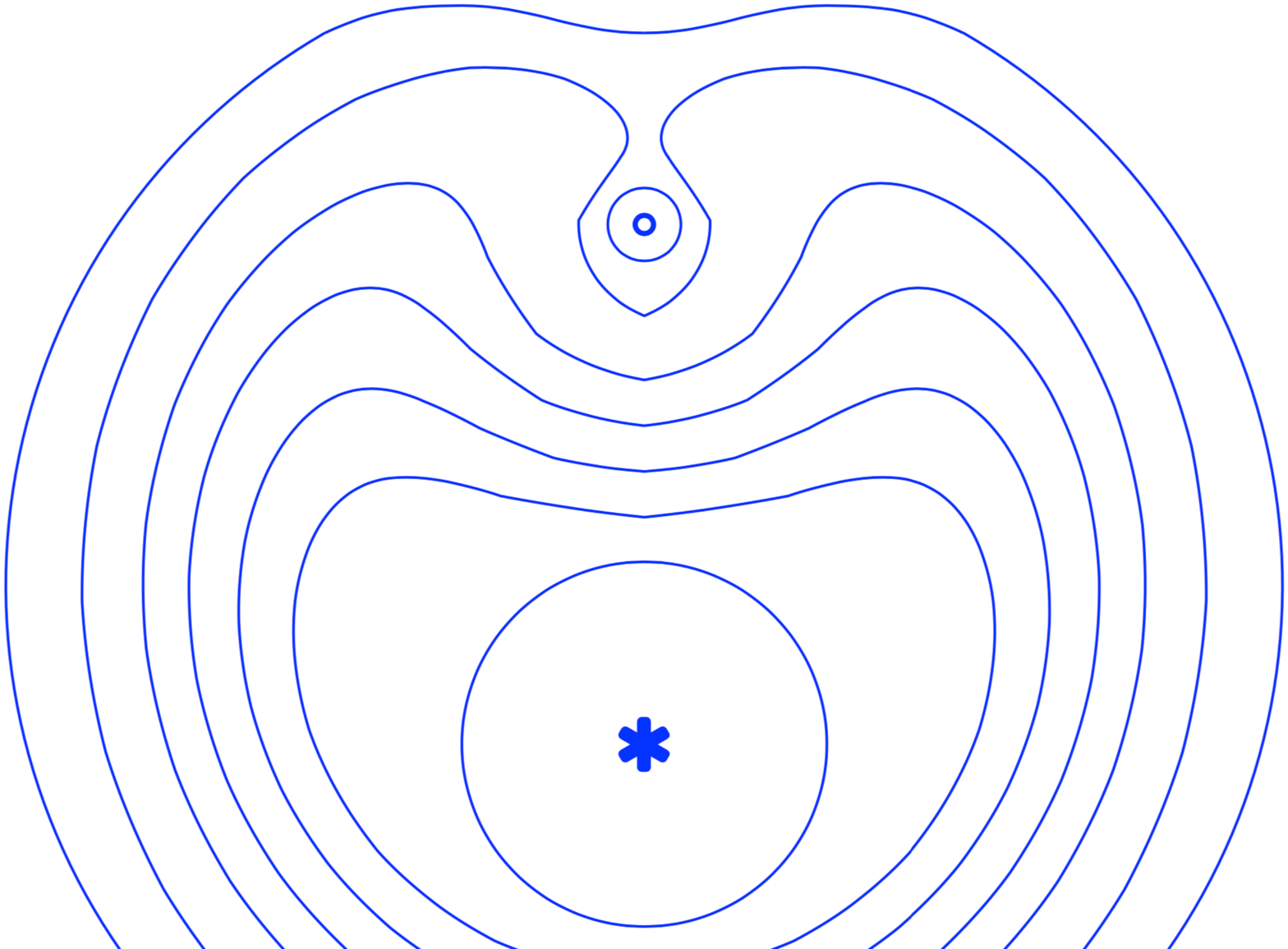}
        \caption{Example of the Stokes phenomenon at half-filling if there are only two auxiliary fields. We display the isolines of the action for the case when the relevant saddle point (local minima, denoted by the star) is accompanied by the irrelevant one and the zero of determinant (top part of the plot, denoted by the open circle).}
        \label{fig:stokes_2D}
\end{figure}

These single and multi-blob configurations have consequences for the fermions, as we attempt to illustrate in Fig.~\ref{fig:charge}. We first define the equal-time fermion Green's function in position-time representation
\begin{equation}
    \label{fermion_propagator}
    g(x, y, \tau) = - \vev{\hat{a}_x(\tau) \hat{a}^{\dagger}_y(\tau) },
\end{equation}
where we have written the expression for particles and an analogous expression exists for the holes. We compute this expression on a given saddle point configuration, for fixed spatial positions $x$ and $y$ as a function of $\tau$. This quantity forms a closed curve in the complex plane due to periodic boundary conditions for the auxiliary fields. Furthermore, for certain locations of the source and sink, this curve exhibits a non-trivial winding around the origin in the complex plane. We define the winding number of the propagator for a given source and sink location as follows
\begin{eqnarray}
    W(x,y) &\equiv& \frac{1}{2\pi i} \oint_{\gamma} \frac{dz}{z}\\
    &=& \frac{1}{2\pi i} \int^{\beta}_0 \frac{1}{g(x, y, \tau)} \frac{\partial g(x,y, \tau)}{\partial \tau} d\tau,\nonumber
    \label{fermion_winding}
\end{eqnarray}
where in the first equality we have used $z \equiv g(x,y,\tau)$ and $\gamma$ refers to the closed curve swept out by the propagator in the complex plane. For the one-blob configuration in Fig.~\ref{fig:charge}, we have plotted the Green's function contour for two different sinks, with the source fixed at the center of the blob. In Fig.~\ref{fig:charge}\textcolor{red}{-1(b)}, the sink is located on the opposite sublattice of the source and shows a non-trivial winding number of $+1$, while in Fig.~\ref{fig:charge}\textcolor{red}{-1(c)} the sink is located on the same sublattice of the source and shows a trivial winding of $0$. We thus see that there exists a correlation between fermion winding number, saddle points, and sublattice symmetry.

We have observed that, for the multi-blob configurations, blobs with the same sign lie on the same sublattice while blobs with opposite signs lie on opposite sublattices. The latter is depicted in Fig.~\ref{fig:charge}\textcolor{red}{-2(a)} where we have a configuration containing a blob-anti-blob pair, and in Fig.~\ref{fig:charge}\textcolor{red}{-2(b)} and~\ref{fig:charge}\textcolor{red}{-2(c)} we observe the same correlation between sublattice symmetry and fermion winding number that was observed for the one-blob configuration. However, in ~\ref{fig:charge}\textcolor{red}{-2(d)}, we see a non-trivial winding number of $+2$ where the sink and source were taken to be the centers of the two blobs. A two-blob configuration is depicted in~\ref{fig:charge}\textcolor{red}{-3(a)}, where again, the winding number is trivial for source and sink on the same sublattice (Fig.~\ref{fig:charge}\textcolor{red}{-3(b)}). The winding number is non-trivial and equal to $-1$ for source and sink on different sublattices (Fig.~\ref{fig:charge}\textcolor{red}{-3(c)}). We note that winding number $\pm 2$ was not observed for the two-blob configuration. We assume that a similar correlation exists between the winding number and the construction of saddle point configurations with a larger number of (anti-)blobs, and thus the winding number can be used for the classification of saddle points. However, we have left the detailed study of this point to future work.

\begin{figure}
        \centering
        \includegraphics[width=3.5in, angle=0]{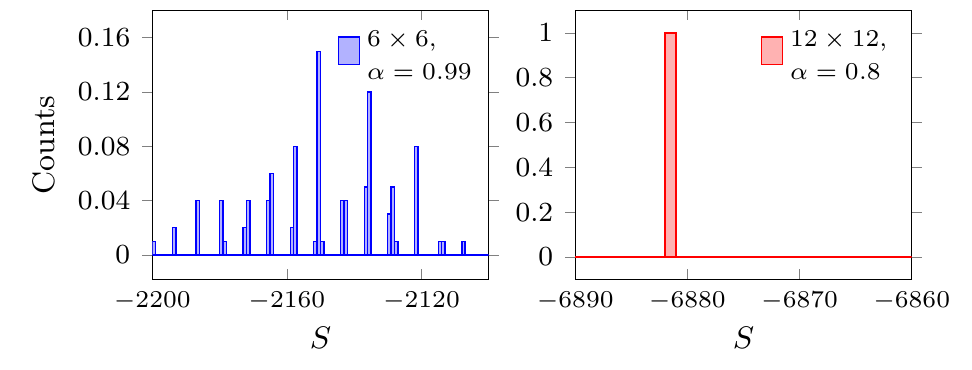}
        \caption{(LHS) The distribution of the action of saddle point configurations at half filling for $\alpha=0.99$. Results are shown for $6\times6$ lattice with $N_{\tau}=256$ and $\beta=20.0$, $U=3.8 \kappa$. The situation again becomes substantially worse: non-vacuum saddle points play a significant role. (RHS) Quite the opposite situation is observed at $\alpha=0.8$. In this case, we see only one saddle point even in the case of larger, $12\times12$ lattice with the same $N_{\tau}$, $\beta$ and $U$.}
        \label{fig:large_and_optimal_alpha_histogram}
\end{figure}

One expects that the dependence of the thimbles decomposition on the Hubbard coupling should reflect the changing physics in the strongly-coupled phase. The dependence of the real part of the action of the various saddles on the coupling $U$ at half-filling is shown in Fig.~\ref{fig:history_U} for the case where the charge-coupled Hubbard field dominates ($\alpha=0.9$). In order to track the location of the saddles in a continuous manner we have used the GF in the downwards direction after small shifts of the on-site interaction $U$. This means that we start from saddle points at large $U$, then slightly decrease $U \rightarrow U-\delta U$ and search for the new locations of the local minima by starting GF from the old saddles. This procedure is repeated to cover the desired interval of $U$. We have found that sometimes the profiles obtained in this way experience sharp decays into the vacuum saddle. This behavior implies that the corresponding saddle point becomes irrelevant.

\begin{figure}
        \centering
        \includegraphics[scale=0.10, angle=0]{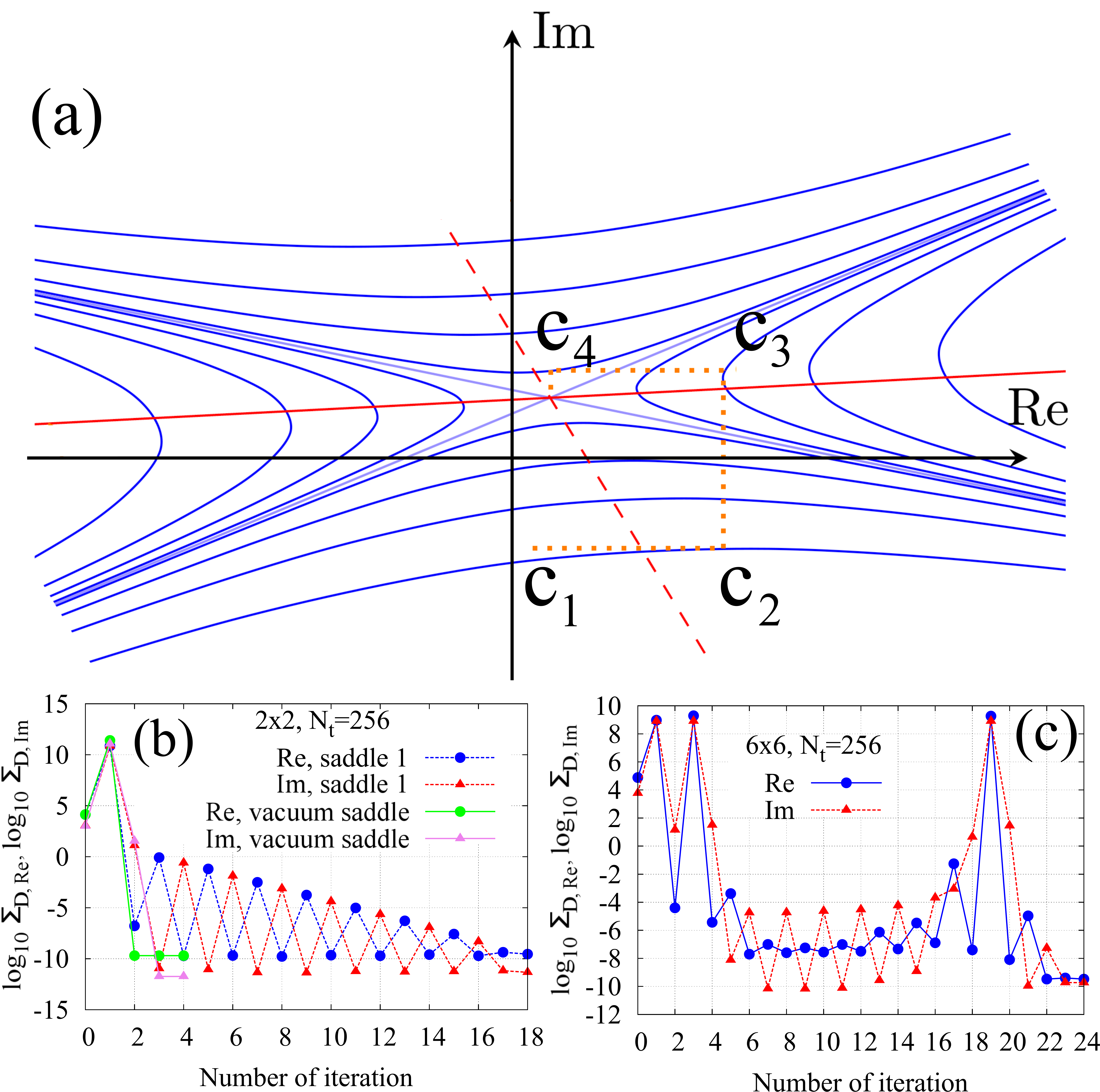}
        \caption{(a) General schematic illustration of the algorithm which searches for complex saddle points (1D case): $c_1$ is the initial position, segment $c_1-c_2$ corresponds to the downward flow, segment $c_2-c_3$ corresponds to the upward flow and so on. (b) Example of search processes for a $2\times2$ lattice with $N_{\tau}=256$, $\beta=20.0$, $U=2.0 \kappa$, $\mu=\kappa$ and $\alpha=1.0$: shorter process converges to vacuum saddle point and longer one shows convergence to non-vacuum localized saddle point. (c) Example search process for a $6\times6$ lattice with $N_{\tau}=256$, $\beta=20.0$, $U=2.0 \kappa$, $\mu=\kappa$ and $\alpha=0.9$: it illustrates the case when the process collides with a zero of the determinant on the way, seen in the large spikes for both the real and imaginary parts. The y-axis in figures (b) and (c) labels the sum of the squares of the first derivatives of $\mathrm{Re}S$ with respect to the real or imaginary parts of the fields at each site.}
        \label{fig:general_complex_search}
\end{figure}

Before we proceed further, the last point needs to be clarified. Usually, a thimble and its corresponding saddle point are classified as ``relevant'' if their intersection number, $k_\sigma$, is nonzero. However, things can be different if the so-called Stokes phenomenon occurs. This situation implies that several saddles are now connected by one thimble. Here we consider this situation at half-filling, when there is no sign problem and all relevant thimbles and saddle points are confined within $\mathbb{R}^N$. In this case, all eigenvectors of the Hessian matrices, $\Gamma_\sigma$, for saddles located within $\mathbb{R}^N$ have their components either purely real or imaginary. At the local minimum of the action within $\mathbb{R}^N$, which is a relevant saddle point, all real eigenvectors of $\Gamma_\sigma$ correspond to positive eigenvalues. However, it can happen that some $\mathcal{N}_\sigma>0$ real eigenvectors correspond to negative eigenvalues of $\Gamma_\sigma$. This situation is illustrated in Fig.~\ref{fig:stokes_2D}. Because the thimbles attached to local minima cover the entire $\mathbb{R}^N$, saddles which have at least one real eigenvector corresponding to a negative eigenvalue of $\Gamma_\sigma$, do not participate in the sum~(\ref{eq:thimbles_sum_with_phases}), and thus are irrelevant. Simply counting the intersection points is impossible in this case as $\dim (\mathbb{R}^N \cap \mathcal{K}_\sigma )=\mathcal{N}_\sigma>0$ for such saddles.  The decay of a saddle, if we start from a slightly shifted field configuration, means that a negative eigenvalue of $\Gamma_{\sigma}$ with a corresponding real eigenvector has appeared. It then follows that this situation indeed corresponds to the transition between relevant and irrelevant status for the saddle point.

\begin{figure}
        \centering
        \includegraphics[scale=0.14, angle=0]{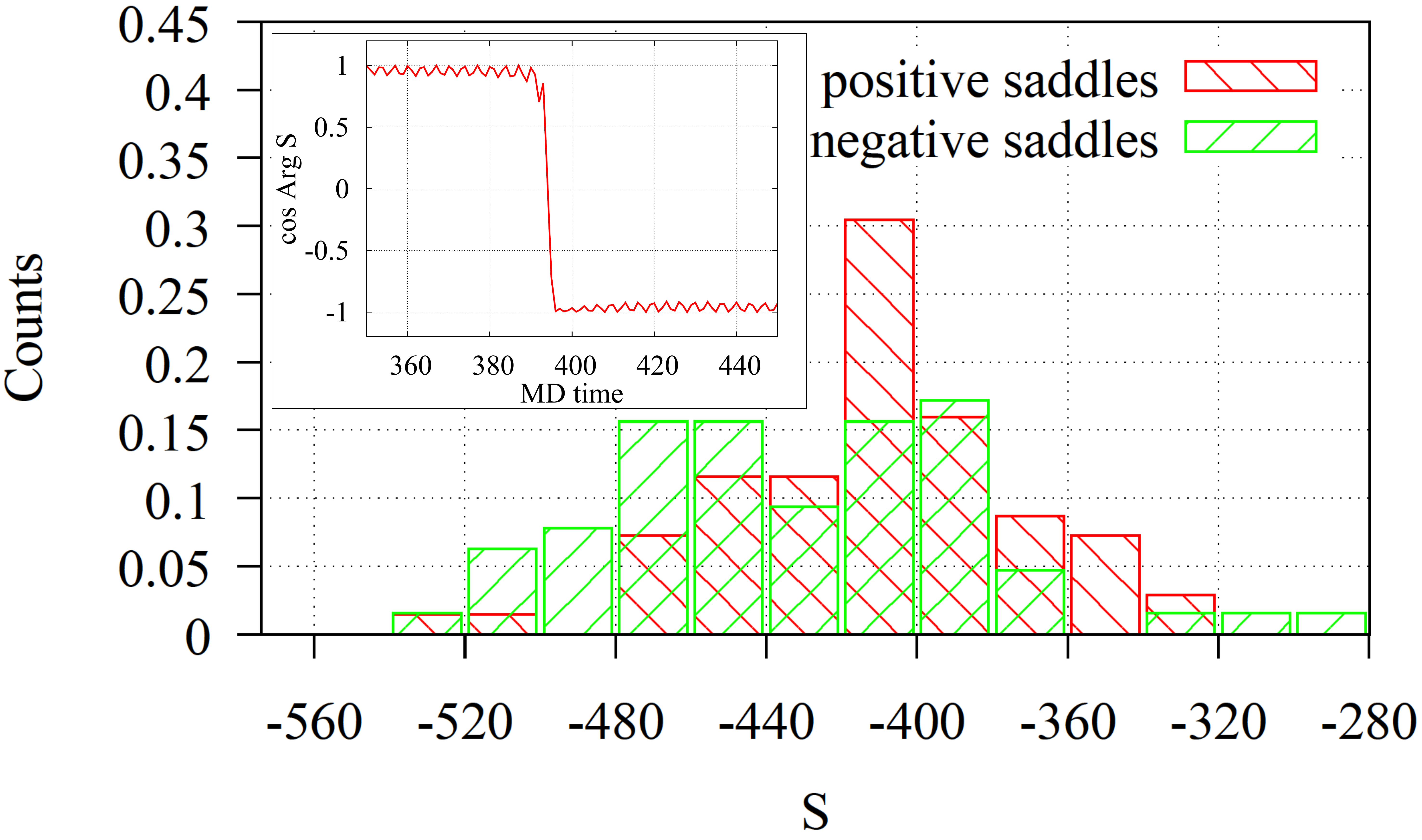}
        \caption{The distribution of the action of saddle point configurations at $\mu=\kappa$ for $\alpha=1.0\times10^{-4}$. Results are shown for a $6\times6$ lattice with $N_{\tau}=256$ and $\beta=20.0$, $U=3.8 \kappa$. Saddle points with positive and negative sign are shown separately in red and green respectively. Inset: History of $\mbox{Im} S$ during an HMC update of the field configuration showing the tunneling between thimbles.}
        \label{fig:small_alpha_finite_mu_histogram}
\end{figure}

Using these ideas we can interpret from Fig.~\ref{fig:history_U} that at small coupling, the trivial vacuum is the only relevant saddle. As we move to larger coupling, multiple non-trivial relevant saddles appear and above $U \approx 3.4 \kappa$, we see two of them which are evenly spaced. Thus, we should expect that, at fixed $\alpha$ and large $U$, more and more non-trivial multi-blob saddles become relevant once we approach AFM phase. This interpretation is also supported by the previous histograms cf. Fig.~\ref{fig:intermediate_alpha_histogram}.

 Finally, in Fig.~\ref{fig:large_and_optimal_alpha_histogram}\textcolor{red}{(a)} we show how the situation becomes worse as we further increase the parameter $\alpha$, thus suggesting that there exists a ``sweet spot" which possesses an advantageous structure for the thimbles decomposition. This regime is illustrated in Fig.~\ref{fig:large_and_optimal_alpha_histogram}\textcolor{red}{(b)}, where even for large lattices ($12\times12$) at $U=3.8 \kappa$ (which corresponds to the AFM phase transition), only the vacuum saddle contributes at $\alpha=0.8$. One can compare this situation with that depicted in the lower panel middle plot of Fig.~\ref{fig:intermediate_alpha_histogram}. A more detailed study of this regime is made below, accompanied by the study of saddles points away from half-filling.

\begin{figure}[]
   \centering
   \subfigure%
             {\label{fig:all_schemes_thimbles_1a}\includegraphics[width=0.4\textwidth,clip]{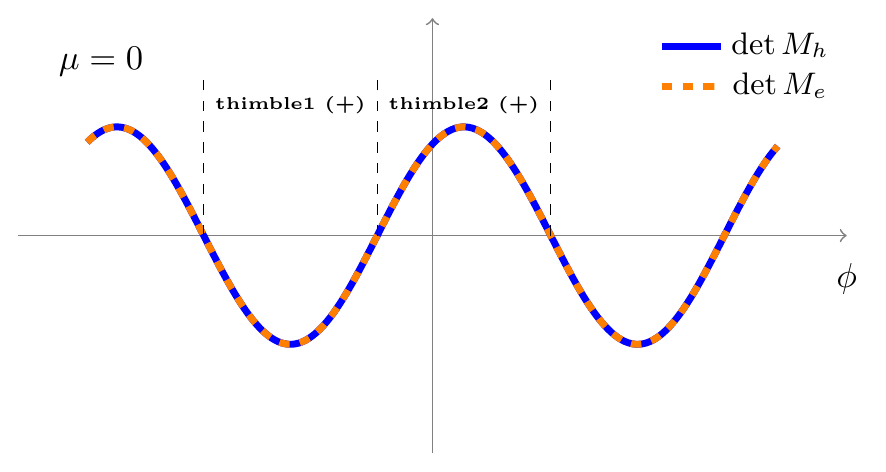}}\vfill
   \subfigure%
             {\label{fig:all_schemes_thimbles_1b}\includegraphics[width=0.4\textwidth,clip]{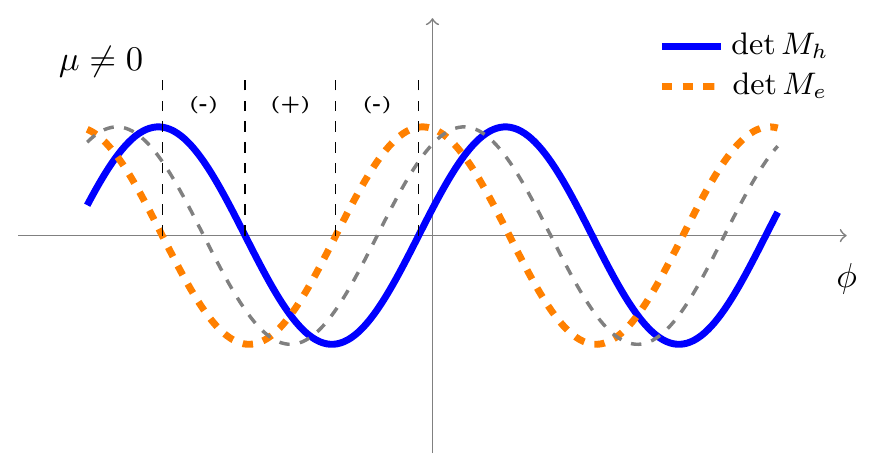}}
      \caption{Schematic diagrams which explain the appearance of ``negative'' thimbles at nonzero chemical potential in the case when only the spin-coupled field is present ($\alpha=0.0$)}
   \label{fig:schemes_one_field}
\end{figure}

\subsection{\label{subsec:spstudy:nonzero_mu}Saddle points at nonzero chemical potential}

Away from half-filling one can not rely on the naive application of the GF equations in order to find the saddle points. This is due to the fact that the downward GF ends up on a saddle point only if the initial configuration was exactly on the corresponding thimble. Since we can not generate  those configurations (at least without prior knowledge about the saddle points), another method should be employed. We use a procedure similar to Powell's method to search for local minima. The algorithm is illustrated schematically in  Fig.~\ref{fig:general_complex_search}\textcolor{red}{(a)}, for a single complex field. The minimization procedure consists of alternating GF steps for constant imaginary and real parts of the field. The even iterations consist of GF in the downward direction with fixed $\mbox{Re} \Phi_j=\Phi^{(R)}_j$, where $\Phi_j \equiv \Phi^{(R)}_j + i\Phi^{(I)}_j$ represents both complex auxiliary fields. The flow stops when it reaches the local minimum. The odd iterations consist of upward GF with fixed $\mbox{Im} \Phi_j=\Phi^{(I)}_j$ and terminate when a local maximum or zero of determinant has been reached, where $\mbox{Re} S \rightarrow \infty$. The convergence can be controlled by monitoring the quantity, $\Sigma_{D, Re/Im} \equiv \sum_i | \partial \mbox{Re} S/ \partial \Phi^{(R/I)}_i |^2$ (with the sum running over all sites in the spatial and temporal directions) after each iteration, with $\Sigma_{D, Re}$ reaching the level of numerical precision (typically $10^{-10}$) during even iterations and $\Sigma_{D, Im}$ during odd iterations (assuming the flow did not collide with a zero of determinant). Some examples are shown in Fig.~\ref{fig:general_complex_search}\textcolor{red}{(b)} and Fig.~\ref{fig:general_complex_search}\textcolor{red}{(c)}. In the former, one can see two examples of the iterations on a $2\times2$ lattice. Here we see that one converges into the vacuum saddle, which is uniformly shifted into the complex plane ($\mbox{Re} \phi_{x,\tau}=\mbox{Re} \chi_{x,\tau}=\mbox{Im} \chi_{x,\tau}=0$, $\mbox{Im} \phi_{x,\tau}=\phi_0$), while the other converges into a non-trivial saddle, which is non-uniform both in space and Euclidean time. The latter figure demonstrates an example for a $6\times6$ lattice, where the iterations collided with a zero of the determinant on the way, but nevertheless converged afterwards. 

Away from half-filling, the initial configurations were generated using a phase quenched HMC, using the algorithms already described in~\cite{Conformal_PhysRevB.99.205434}. Thus, only the absolute value of $\ln \det (M_{\text{el.}} M_{\text{h.}})$ was taken into account during the Monte Carlo procedure. Usually, the initial configurations are generated along some contour in $\mathbb{C}^N$, uniformly shifted from $\mathbb{R}^N$, in order to approach the thimble. This is not surprising as we have found that this constant shift into complex space applies to the vacuum saddle at $\mu \neq 0$. The procedure of using a constant shift was performed at $\alpha=0.8$ and $\alpha=0.9$, where the charge-coupled field dominates. If $\alpha=0$, the thimbles and saddles again lie within $\mathbb{R}^N$, since both fermionic determinants are real. However, as discussed previously, this property of the fermionic determinants leads to a loss of ergodicity for HMC. Thus, in order to explore the case when the spin-coupled field dominates, we use small $\alpha=10^{-4}$ and generate configurations without a shift into the complex plane. Even such a small, nonzero value of $\alpha$ is enough to restore ergodicity, as one can see in the inset in Fig.~\ref{fig:small_alpha_finite_mu_histogram}. This inset shows the history of $\arg S$ during one trajectory in HMC update. If $\alpha=0$, all thimbles have $\cos \arg S=\pm 1$ again due to the fact that $\det (M_{\text{el.}} M_{\text{h.}}) \in \mathbb{R}$. Thimbles with different signs are separated by zeros of the determinants, since they are branch points of the logarithm. Here we have a small but nonzero $\alpha$, and thus the $\cos \arg S$ only approaches  $\pm 1$. A sharp transition is observed in the inset in Fig.~\ref{fig:small_alpha_finite_mu_histogram} which shows us that the algorithm still can tunnel between different thimbles. This tunneling was, in fact, quite frequent and was observed in more than half of the Monte Carlo updates. This is a further confirmation that the HMC is ergodic.

\begin{figure}
        \centering
        \includegraphics[width=3.5in, angle=0]{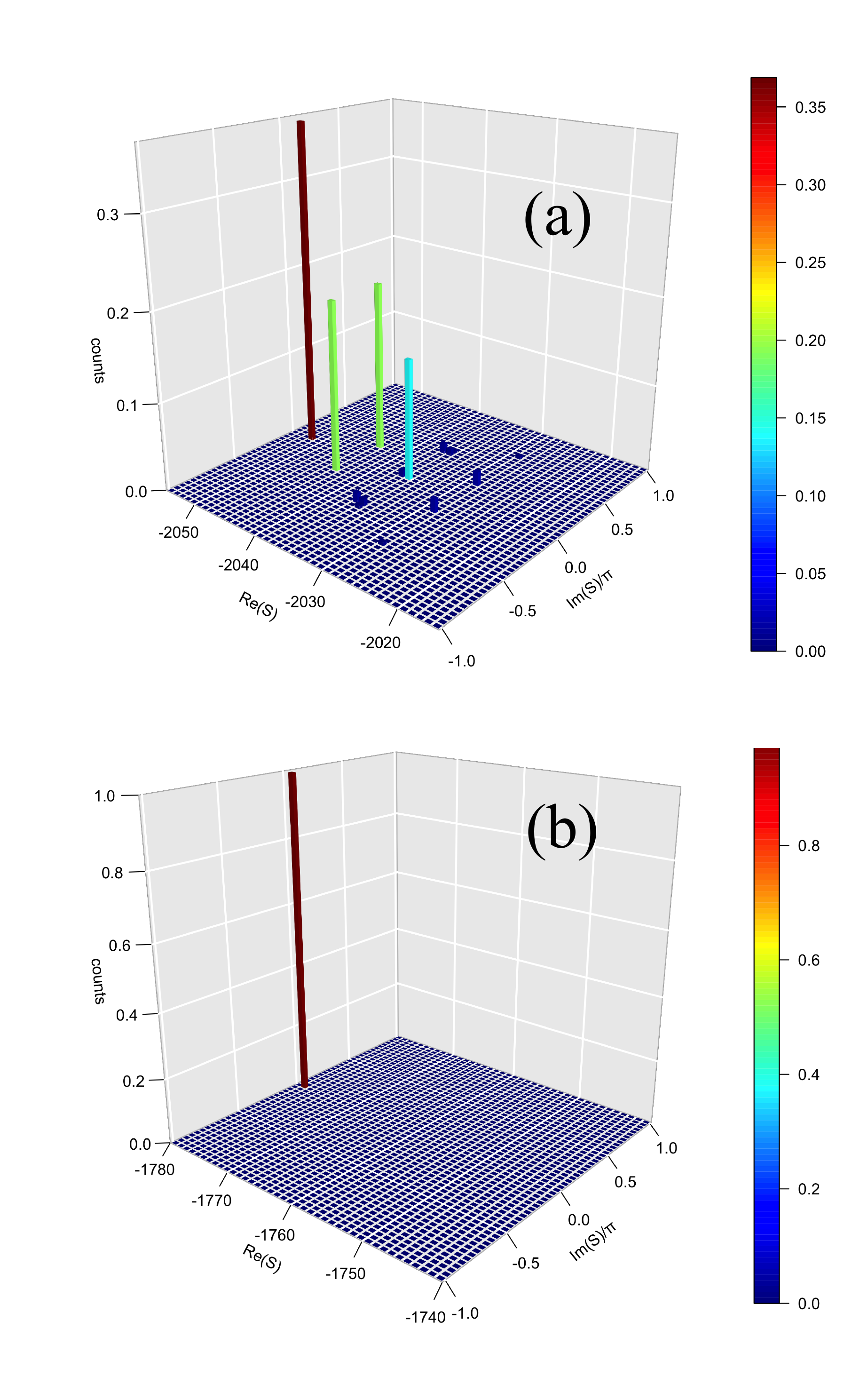}
        \caption{The distribution of the action of saddle point configurations at $\mu=\kappa$ for $\alpha=0.9$ (a); and $\alpha=0.8$ (b). Results are shown for a $6\times6$ lattice with $N_{\tau}=256$ and $\beta=20.0$, $U=3.8 \kappa$. As the action is complex away from half-filling, the histogram is plotted simultaneously both for real and imaginary parts of the action. The set of saddle points is similar to the results at half-filling at the same $\alpha$ (see Fig.~\ref{fig:intermediate_alpha_histogram}). Plot (b) shows that again, only one (shifted trivial vacuum) saddle point can be found for $\alpha=0.8$.}
        \label{fig:large_alpha_finite_mu_histogram}
\end{figure}

\begin{figure}
        \centering
        \includegraphics[scale=0.31, angle=0]{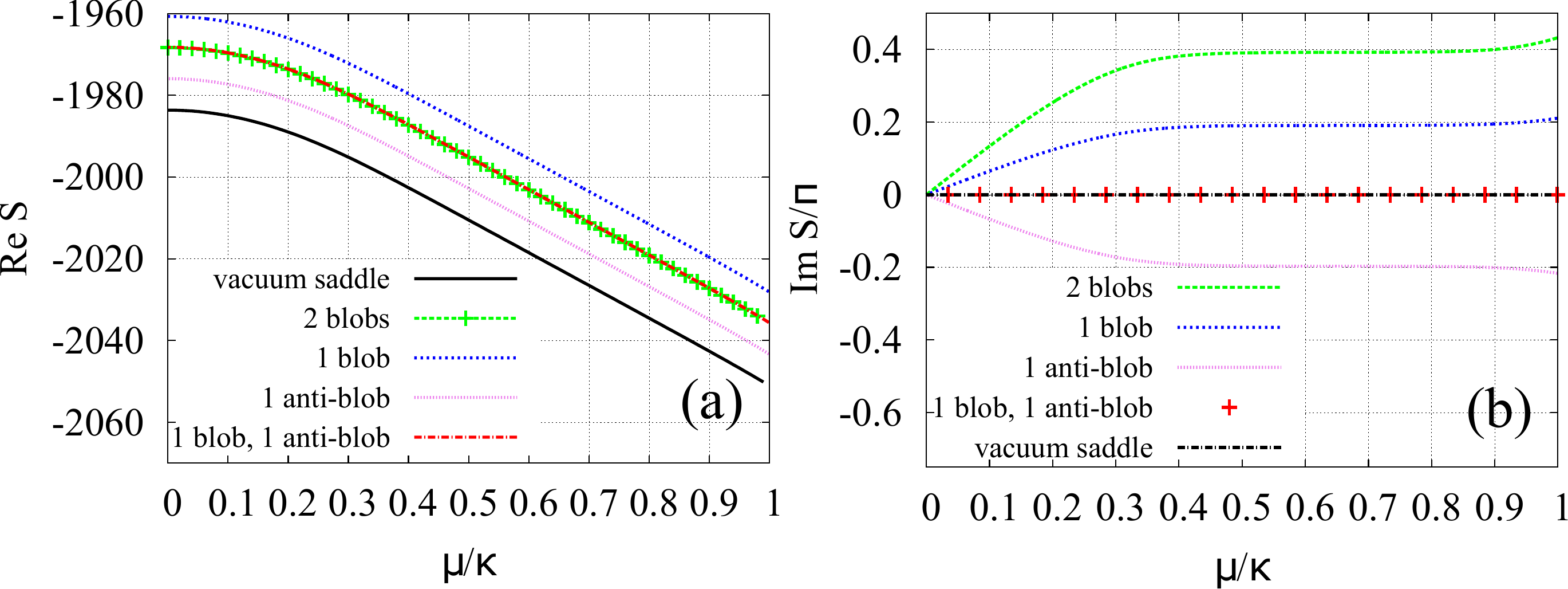}
        \caption{Evolution of saddle points at $\alpha=0.9$ with increasing chemical potential. Real (a) and imaginary (b) parts of the action are shown. Saddle points remain roughly equally spaced in terms of the real part of their action, while their phases start to diverge with increasing $\mu$. Results are shown for a $6\times6$ lattice with $N_{\tau}=256$ and $\beta=20.0$, $U=3.8 \kappa$.}
        \label{fig:history_mu}
\end{figure}

Another concern regarding our GF procedure is the question of convergence of the alternating iterations. Unfortunately, the procedure we have used does not converge for an arbitrary saddle. The criterion for the convergence of the procedure can be derived from the fact that the distance to the saddle point should decrease after a full round of four iterations. 
The exact formulation can be expressed in terms of the Hessian matrix
$\Gamma = \begin{pmatrix} 
A & C \\
C & B 
\end{pmatrix}$.
The Hessian is written in terms of $2N_s N_{\tau} \times 2N_s N_{\tau}$ blocks $A_{i,j} \equiv \partial^2 \mbox{Re} S/\partial \Phi^{(R)}_i \partial \Phi^{(R)}_j$, $B_{i,j} \equiv \partial^2 \mbox{Re} S/\partial \Phi^{(I)}_i \partial \Phi^{(I)}_j$, and $C_{i,j} \equiv \partial^2 \mbox{Re} S/\partial \Phi^{(R)}_i \partial \Phi^{(I)}_j$. Using these matrices, the minimization procedure is guaranteed to converge if both $A$ and $-B$ are positive-definite, and each of the eigenvalues, $\lambda_i$, of the matrix $A^{-1} C B^{-1} C$, which characterizes the update of the fields after two subsequent iterations, satisfy 
\begin{equation}
\label{eq:convergence}
  |\lambda_i| < 1.
\end{equation}
The latter is actually a constraint on $|\arg \partial_i \partial_j S|$. If all of the second derivatives are real, $C=0$, and thus $|\lambda_i| = 0$. If $|\arg \partial_i \partial_j S|$ increases, with $A$ and $(-B)$ still remaining positive-definite, the thimble in the vicinity of the saddle point is no longer parallel to $\mathbb{R}^N$, but starts to ``rotate'' in complex space. In the $1$D case illustrated in Fig. ~\ref{fig:general_complex_search}\textcolor{red}{(a)}, $|\lambda| < 1$ simply means that $|\arg \partial^2 S |_{z_\sigma}|<\pi/4$.

We again start from small $\alpha$, which corresponds to a dominant spin-coupled field. In this case, all saddle points are located at $\phi_{x,\tau}=0$ and $\mbox{Im} \chi_{x,\tau}=0$ with their phases $\cos \arg S=\pm 1$. Results are shown in Fig.~\ref{fig:small_alpha_finite_mu_histogram}, separately for positive and negative saddles. In general, we observed a very large variety of saddles with non-uniform structures both in space and time. It is almost impossible to characterize them, since their actions form a quasi-continuum distribution. Furthermore, positive and negative saddles almost compensate each other in this case (see histograms in~\cite{Ulybyshev:2019hfm}), and thus the part of the residual sign problem stemming from the phase factors in eq.~(\ref{eq:thimbles_sum_and_integral}) is quite strong. The qualitative explanation for such behavior can be derived from the schematic illustrations in Fig.~\ref{fig:schemes_one_field}. At half filling, the fermionic determinants for electrons and holes are identical for $\alpha=0.0$, thus the sign problem is absent, but, according to the previously mentioned results (see Fig.~\ref{fig:small_alpha_histogram},~\ref{fig:spin_saddles}) we have many thimbles in $\mathbb{R}^N$, separated by zeros of the determinants. Once the chemical potential shifts from zero, the two determinants are no longer identical, the domain walls between thimbles are split, and ``negative'' thimbles immediately appear along the borders between ``positive'' thimbles. Since we observe a large variety of thimbles at half-filling, the situation can only become worse at $\mu \neq 0$.

Results for large $\alpha$ are shown in Fig.~\ref{fig:large_alpha_finite_mu_histogram}. At $\alpha=0.9$, the distribution of both $\mbox{Re} S$ and $\mbox{Im} S$ show the same characteristic behavior as was observed at half filling (Fig.~\ref{fig:intermediate_alpha_histogram}) with saddle points equidistantly spaced in action. The difference is that the step in the action is now a complex number. More precisely, the properties of the saddle points can be understood from Fig.~\ref{fig:history_mu}, which shows the evolution of the saddle points as one goes away from half-filling. We see a continuous evolution of the same system of blobs and anti-blobs. The difference between them in $\mbox{Re} S$ remains constant (Fig.~\ref{fig:history_mu}\textcolor{red}{(a)}), while $\mbox{Im} S$ increases and blob and anti-blob configurations acquire opposite phases. As obvious from our previous discussion, the general rule for the approximate action of saddle point is now $S_{n_1, n_2} = S_0+ n_1 \Delta S + n_2 \overline{\Delta S}$, where $n_1$ is number of blobs and $n_2$ is the number of anti-blobs in the configuration. Thus, the actions of the saddle points form a triangle in complex space as can be observed in Fig.~\ref{fig:large_alpha_finite_mu_histogram}\textcolor{red}{(a)}.  Another interesting consequence is that not only the vacuum, but also configurations with an equal number of blobs and anti blobs have $\mbox{Im} S=0$, which effectively decreases the complexity of the sign problem.

\begin{figure}
        \centering
        \includegraphics[scale=0.23, angle=0]{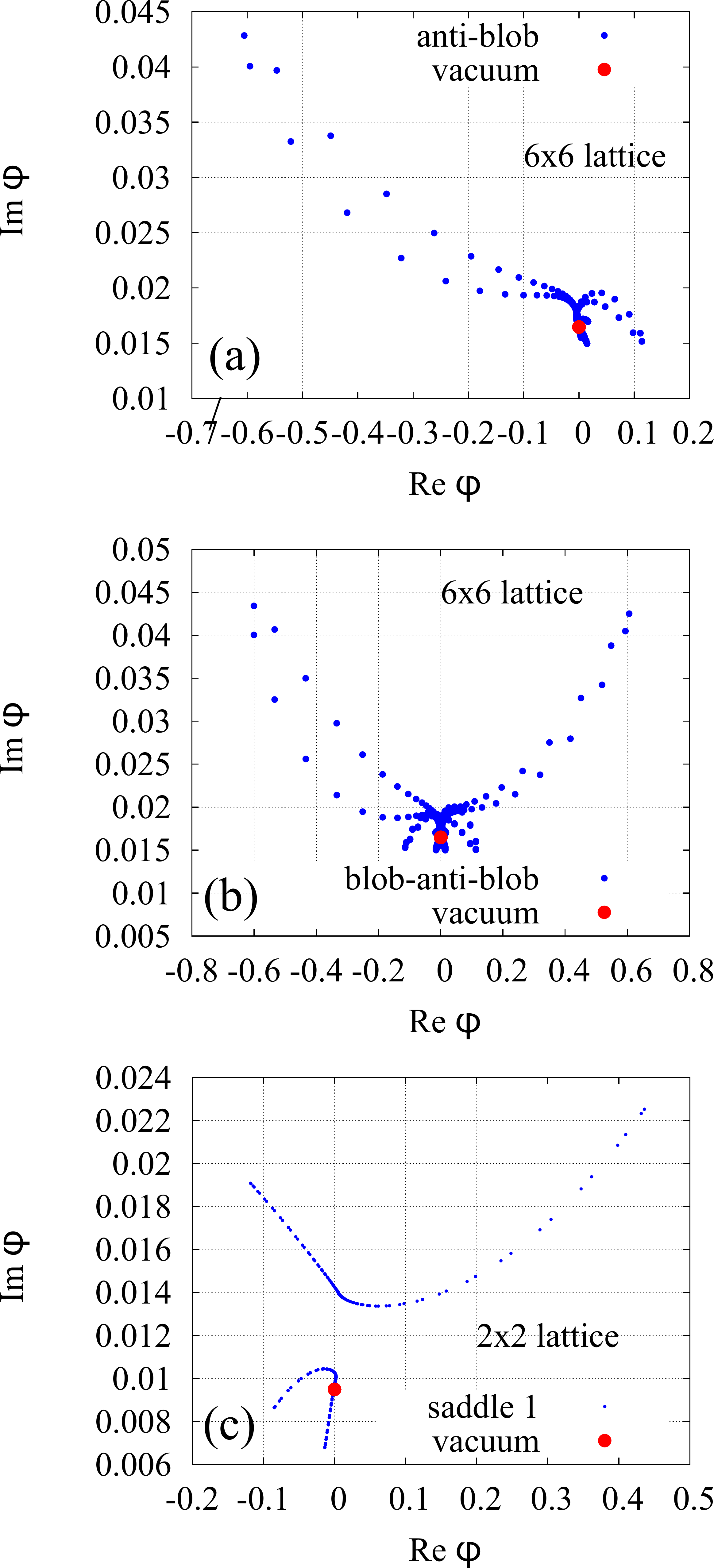}
        \caption{Field configurations at saddle points for mostly charge-coupled auxiliary field at finite chemical potential. (a,b) $6\times6$ lattice with $\alpha=0.9$, $U=3.8 \kappa$ (these plots correspond to the histogram (a) in the Fig.~\ref{fig:large_alpha_finite_mu_histogram}); (c) $2\times2$ lattice with $\alpha=1.0$ and  $U=2.0 \kappa$, displayed here to show how the non-trivial saddle point looks like in the situation where we perform HMC with GF. The other parameters are $N_{\tau}=256$, $\mu=\kappa$, $\beta =20.0$. The $\chi$-field is always equal to zero, and the complex values of all $\phi$-fields are projected onto a single complex plane. The vacuum field configuration corresponds to all $\phi$-fields uniformly shifted into the complex plane along the imaginary axis. The saddle points, which are separated in action from the vacuum, for the $6\times 6$ lattice preserve generally the same localized structure shown in the Fig.~\ref{fig:charge}, with the shifts of the imaginary parts of the fields from the vacuum value following the shift of the real parts.}
        \label{fig:charge_mu}
\end{figure}

As we go to finite chemical potential we can also attempt to visualize the saddle point configurations. Unlike Fig.~\ref{fig:charge}, where the field is real, we plot the configuration in the complex plane with each point representing the value of the field at a given lattice site for  $\alpha=0.9$ (Fig.~\ref{fig:charge_mu}). In these plots, the vacuum configuration is the trivial vacuum $\phi=\chi=0$ with an added constant, volume-independent shift of the imaginary part. In Fig.~\ref{fig:charge_mu}\textcolor{red}{(a)}, we plot the configuration of a single anti-blob. The collection of points which extends furthest away from the vacuum all come from the localized region of space and time surrounding the center of the anti-blob. In Fig.~\ref{fig:charge_mu}\textcolor{red}{(a)}, where we display a blob-anti-blob pair, we see that we have have two such collections of points extending away from the vacuum in opposite directions. Each collection comes from the region of space surrounding the centers of the blob and anti-blob respectively. An illustration of the effect of finite-volume on the non-trivial saddles is depicted in Fig.~\ref{fig:charge_mu}\textcolor{red}{(c)} for a $2 \times 2$ lattice where, for a single blob, the structure of the distribution of the field values is distorted as compared to Fig.~\ref{fig:charge_mu}\textcolor{red}{(a)}.

In the sweet spot regime at $\alpha= 0.8$, we detect again only the vacuum saddle  (Fig.~\ref{fig:large_alpha_finite_mu_histogram}\textcolor{red}{(b)}). In principle, such situation should be very beneficial for the thimbles decomposition, since the fluctuations of $\mbox{Im} S$ can be made arbitrarily small. Also, it should improve the ergodicity of the Monte Carlo process, since the integration manifold is no more divided into disconnected domains. However, we should stress that unlike the $\mu=0$ case, the distributions away from half-filling (Fig.~\ref{fig:small_alpha_finite_mu_histogram} and  Fig.~\ref{fig:large_alpha_finite_mu_histogram}) are exact only for $\alpha=10^{-4}$, since we are quite close to the thimble in this case. For $\alpha=0.8$ and $\alpha=0.9$, the histograms are only approximate as the initial configurations for the iterations approach the thimble, but do not lie exactly on it. Furthermore, ``vertically oriented'' saddles, which do not satisfy the convergence condition (\ref{eq:convergence}) can be missed.  However, subsequent QMC calculations support the conclusion that the regime around $\alpha=0.8$ is indeed better for simulations than $\alpha \rightarrow 1.0$.

\begin{figure}
        \centering
        \includegraphics[scale=0.53, angle=0]{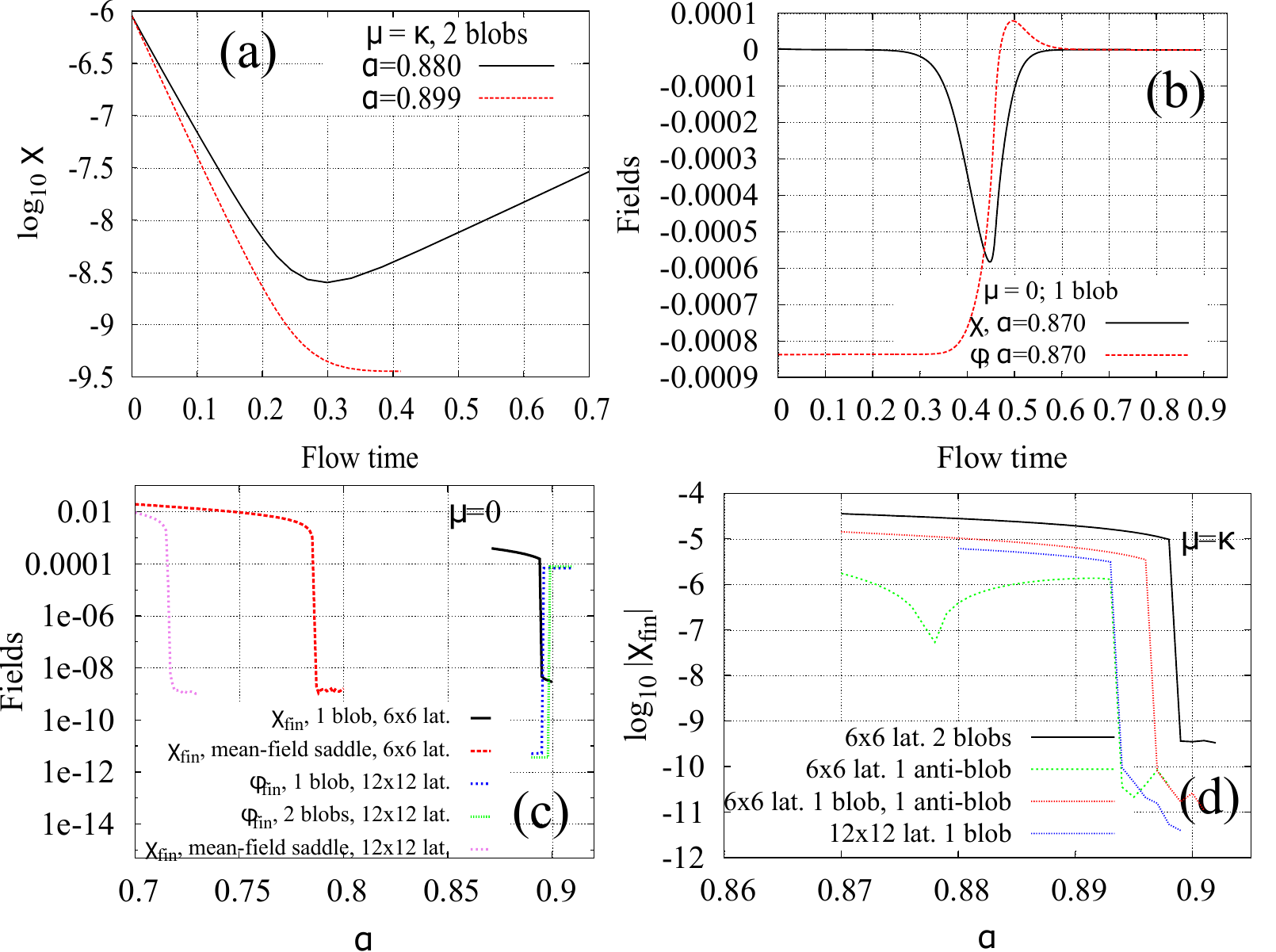}
        \caption{Results of the study of $\alpha$-dependence of saddle points are shown. 
        (a) Example of the $\chi$-flow from the disturbed saddle point in two cases: when the saddle point is relevant and when it is irrelevant. (b) Example of a full flow (both fields vary) originating from the disturbed non-vacuum saddle point when it is irrelevant and the flow ends up in the trivial vacuum.  (c) Summary of results at half-filling. Fields at the end of the flow are shown. The mean-field saddle point for the spin-coupled auxiliary field appears only at $\alpha=0.7...0.8$ while typical non-vacuum saddle points for the charge-coupled field become relevant only at $\alpha\approx0.9$. (d) Decay of non-vacuum saddle points in the case of $\mu=\kappa$. Results are shown for a $6\times6$ lattice with $N_{\tau}=256$ and $\beta=20.0$, $U=3.8 \kappa$.}
        \label{fig:history_alpha}
\end{figure}

\begin{figure}[]
   \centering
   \subfigure%
             {\label{fig:figures/schemes_two_fieldsa}\includegraphics[width=0.45\textwidth,clip]{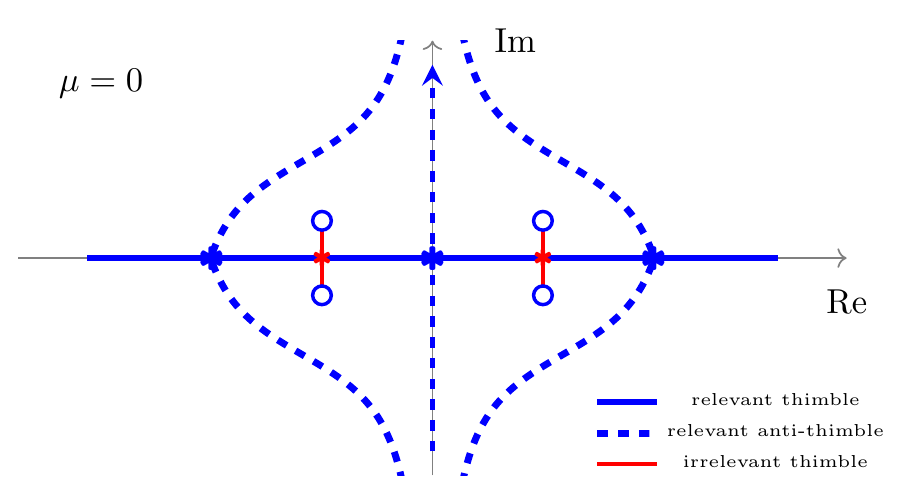}}\vfill
   \subfigure%
             {\label{fig:figures/schemes_two_fieldsb}\includegraphics[width=0.45\textwidth,clip]{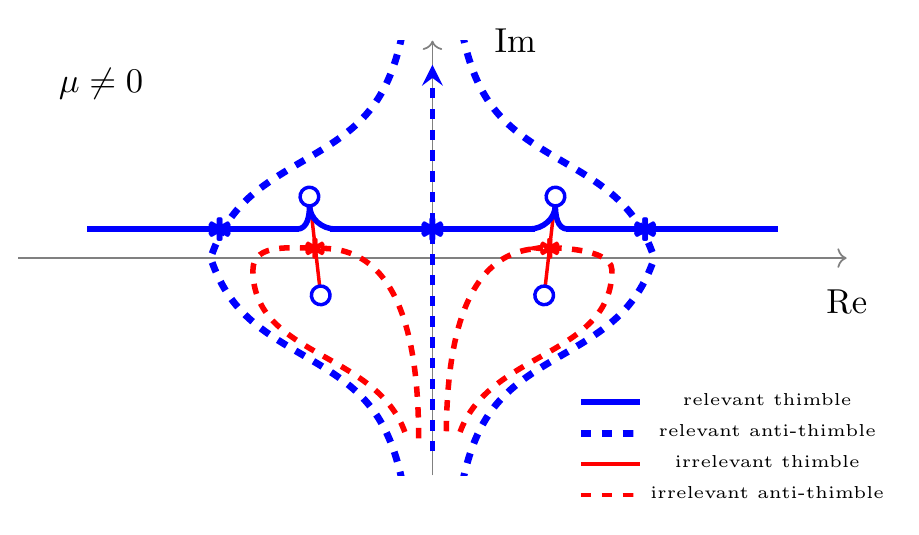}}
   \caption{
   Schematic diagrams which explain the evolution of saddle points and thimbles in the case when both fields are present in the path integral.}
   \label{fig:schemes_two_fields}
\end{figure}

The optimal regime around $\alpha=0.8$ is studied in Fig.~\ref{fig:history_alpha}. We start from half-filling in Fig.~\ref{fig:history_alpha}\textcolor{red}{(c)}, The lower boundary of this region in $\alpha$ corresponds to the  splitting of the vacuum saddle into two mean-field saddles. This splitting is observed by launching GF from a slightly perturbed vacuum (Gaussian noise is added to $\phi_{x,\tau}$ and $\chi_{x,\tau}$). If $\chi_{x,\tau}$ returns to zero, the vacuum is stable, otherwise the final value of $\chi_{x,\tau}$ is non-zero, since the flow arrives at the mean-field saddle point. This is what we see in the $\chi$-profiles for the mean-field saddles in Fig.~\ref{fig:history_alpha}\textcolor{red}{(c)} both for $6\times6$ and $12\times12$ lattices. The jump upwards corresponds to the appearance of the mean-field saddle and marks the lower boundary of the optimal regime. The upper boundary is determined by the decay of the nontrivial saddles into vacuum, analogous to what was observed in Fig.~\ref{fig:history_U}. We use the symmetry, $S(\phi_{x,\tau}, \chi_{x,\tau}) = S(\phi_{x,\tau}, - \chi_{x,\tau})$, and the fact that the saddles are located at $\chi_{x,\tau}=0$ for large $\alpha$. The Hessian matrix is block-diagonal in this case as $\partial^2 S / \partial \chi_{x,\tau} \partial \phi_{x,\tau}=0$. Because it is enough to find at least one instability (negative eigenvalue of $\Gamma$, with real eigenvector), we can study the relevance of saddles separately for $\phi$- and $\chi$-directions. It can be done in two different ways, which should lead to identical results. The first is the calculation fully analogous to the one made for Fig.~\ref{fig:history_U}: we start from large $\alpha$, slightly decrease it and launch GF in downwards direction for both fields, $\phi_{x,\tau}$ and $\chi_{x,\tau}$. At sufficiently small $\alpha$ we see the decay of the saddle into vacuum (example of the flow history for the fields at one particular site is shown in Fig.~\ref{fig:history_alpha}\textcolor{red}{(b)}). Thus, we can plot the dependence of $\phi_{x,\tau}$ after the flow on $\alpha$ and a sharp drop to zero will mark the transition of a non-trivial saddle into an irrelevant one. This approach, however, does not allow us to understand, in which block of the Hessian matrix does the negative, ``unstable'' eigenvalue appear.  Alternatively, one can first use GF, restricted to the $\phi$ fields in order to find the saddle after a small shift of $\alpha$. No instability was found in this case, and the non-trivial saddles can be found for all values of $\alpha$. Finally, we use these saddles, add noise to the $\chi$ fields, and launch the GF, restricted to $\chi$ fields for these configurations. In this case, the instability manifests itself in a finite value of the $\chi$ fields after the flow is performed (the configuration does not return to the initial saddle located at $\chi_{x,\tau}=0$). Thus, we can plot $\chi$ field after the flow and the sudden appearance of a nonzero value signals the transition of the non-trivial, relevant saddle into an irrelevant one. Both approaches are demonstrated in Fig.~\ref{fig:history_alpha}\textcolor{red}{(c)} for $6\times 6$ and $12\times 12$ lattice.  At $\alpha \approx 0.89$, the final value of $\chi$ after the flow for $\chi$ fields jumps upwards. Simultaneously, the final value of $\phi$ fields in the full flow goes down to the level of numerical errors (typically around $10^{-10}$). This signals an instability in the $\chi$-channel, and thus the non-trivial saddle becomes irrelevant. Remarkably, the results depend neither on the type of saddle point nor on lattice size. We attribute this property to the localized nature of non-trivial saddle points at large $\alpha$. An important observation is that the width of the ''optimal regime'' grows with increasing system size, since the lower border shifts to smaller $\alpha$ for the $12 \times 12$ lattice. This lends strong support to the existence of this regime in the thermodynamic limit.

\begin{figure}[]
   \centering
   \subfigure[]%
             {\label{fig:schemes_hmc_flowa}\includegraphics[trim=0 50 0 0,width=0.45\textwidth,clip]{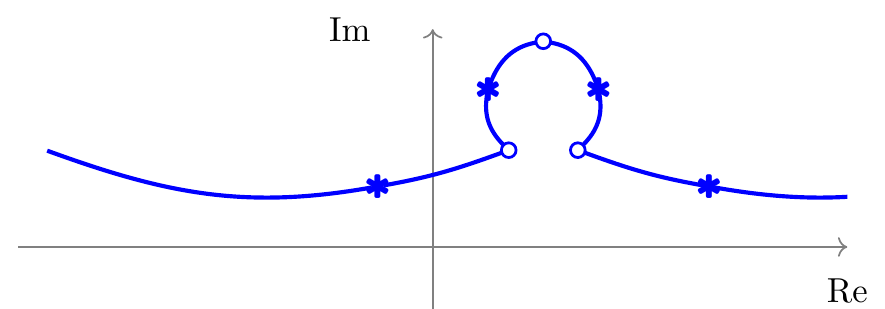}}\vfill
   \subfigure[]%
             {\label{fig:schemes_hmc_flowb}\includegraphics[width=0.45\textwidth,clip]{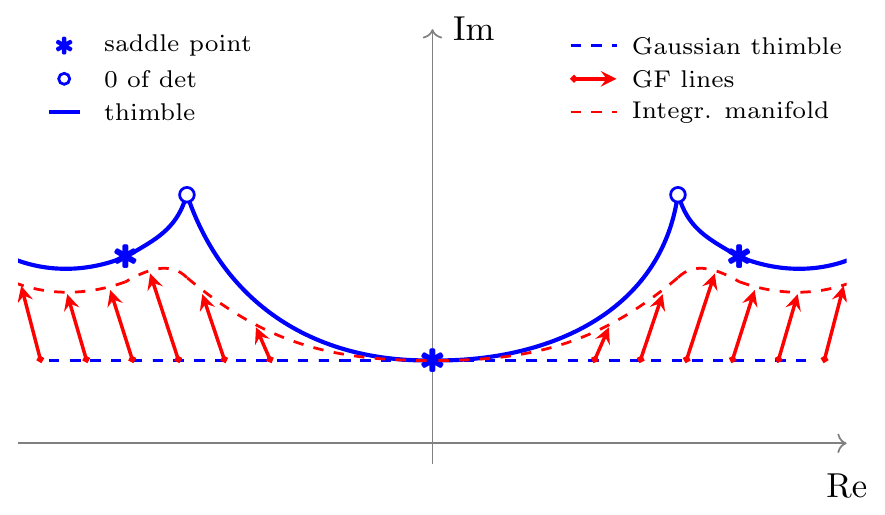}}
   \caption{~\protect\subref{fig:schemes_hmc_flowa} Illustration of ``vertically'' oriented saddle points and thimbles which are lost in the search for complex saddle points. Results are shown for a $2\times2$ lattice with $N_{\tau}=256$ and $\beta=20.0$, $U=2.0 \kappa$, $\mu=\kappa$. ~\protect\subref{fig:schemes_hmc_flowb} Schematic illustration of HMC with gradient flow.}
   \label{fig:schemes_hmc_flow}
\end{figure}

A similar set of calculations was performed for $\mu=\kappa$, where we have used GF restricted to $\mbox{Re} \chi$. The plot in  Fig.~\ref{fig:history_alpha}\textcolor{red}{(a)} demonstrates how the flow switches from the stable regime to eventual decay. We essentially observe the same behavior for the non-vacuum saddles at large $\alpha$ (Fig.~\ref{fig:history_alpha}\textcolor{red}{(d)}): all non-trivial field configurations are unstable in the $\mbox{Re} \chi$ direction if $\alpha<0.89$.  This suggests that at $\alpha>0.89$, the non-vacuum saddles shift into complex space with increasing $\mu$, remaining relevant, while at $\alpha<0.89$ they begin from irrelevant ones at $\mu=0$ and move into complex space for $\mu\neq0$, remaining irrelevant. These situations are depicted in the schematic drawings in Fig.~\ref{fig:schemes_two_fields}, where both relevant and irrelevant saddle points are shown at half-filling and at $\mu \neq 0$.  Another possibility is that the saddles acquire a more ``vertical'' orientation with decreasing $\alpha$ (as shown in Fig.~\ref{fig:schemes_hmc_flowa}). It can happen that the GF along $\mbox{Re} \chi$ can take us away from zero also in this case.  However, there are additional arguments against this based on the results of our HMC over manifolds in complex space at different $\alpha$ described in the next section.

\section{HMC with gradient flow}

In order to check the conclusions concerning the structure of the thimbles decomposition, made in the previous section, we performed several Monte Carlo calculations, with the integration manifolds shifted in complex space towards the thimbles. Following~\cite{Alexandru:2016ejd}, the sequence of deformations of the integration contour can be summarized by
\begin{equation}
\mathcal{Z}=\int_{\mathbb{R}^N} \mathcal{D} \Phi e^{-S[\Phi + i \Phi_0]} = \int_{\mathbb{R}^N} \mathcal{D} \Phi e^{-S[\tilde \Phi]} \det J.
\label{eq:HMC_flow}
\end{equation}
The general idea behind these deformations is shown in the schematic illustration of  Fig.~\ref{fig:schemes_hmc_flowb}. 
First, we perform a uniform shift into the complex space, but only for the charge-coupled field: $\phi \rightarrow \phi+i\phi_0$. We work at large $\alpha$, and this uniform shift moves onto the thimble attached to the vacuum saddle (see Fig.~\ref{fig:charge_mu}) in the Gaussian approximation. Below we will denote this shift as $\Phi \rightarrow \Phi + i \, \Phi_0$. A further shift is made using the GF equations. The quantity $\tilde \Phi \in \mathbb{C}^N$ is the result of the evolution of the field determined by (\ref{eq:flow}), starting from the Gaussian thimble $\Phi + i \, \Phi_0, \Phi \in \mathbb{R}$ with flow time $\mathcal{T}$. We should stress that this algorithm does not follow thimble exactly, it only approaches it in the limit of infinitely large flow time. Thus, the third source of the residual sign problem, the residual fluctuations of $\mbox{Im} S$ appear in calculations. The complex-valued Jacobian of the transformation, $J=\mathcal{D} \tilde \Phi / \mathcal{D} \Phi$, also appears in the integral at this stage. The flow time plays a dual role in this transformation. First, it defines, how close we can approach the thimble, and thus it regulates the residual fluctuations of $\mbox{Im} S$. Second, if the flow time is too large, the flow lines can reach zeros of the determinant, which separate thimbles (see Fig.~\ref{fig:schemes_hmc_flowb}). In this case, the integration domain for the $\Phi$ fields is again split into separate regions and as a result the Monte Carlo process can hardly be expected to be ergodic. And finally, another contribution to the residual sign problem comes from the Jacobian, especially in the case of ``vertically'' oriented thimbles, as shown in Fig.~\ref{fig:schemes_hmc_flowa}.

\begin{figure}
        \centering
        \includegraphics[scale=0.19, angle=0]{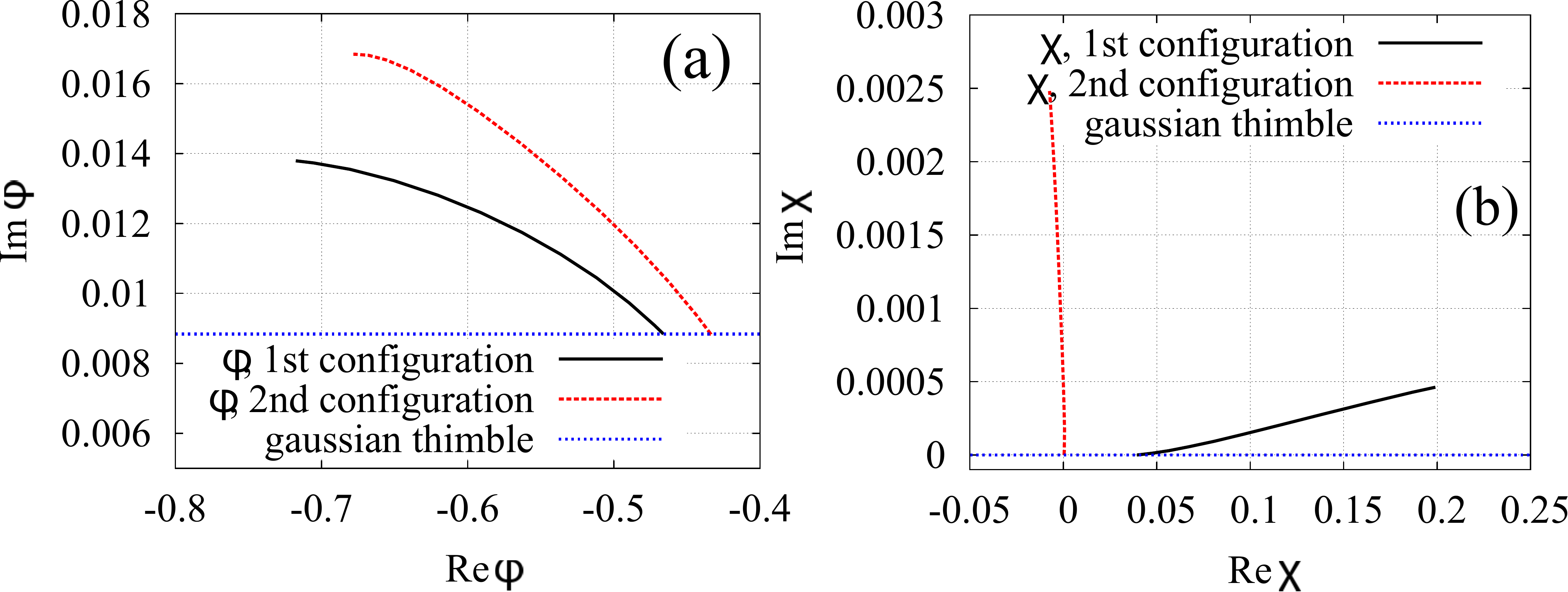}
        \caption{An example of flow profiles for $\phi$ (a) and $\chi$ (b) fields at a single lattice site is shown. The flow starts at the Gaussian thimble attached to the shifted trivial vacuum. }
        \label{fig:example_hmc_flow}
\end{figure}

\begin{figure}
        \centering
        \includegraphics[scale=0.18, angle=0]{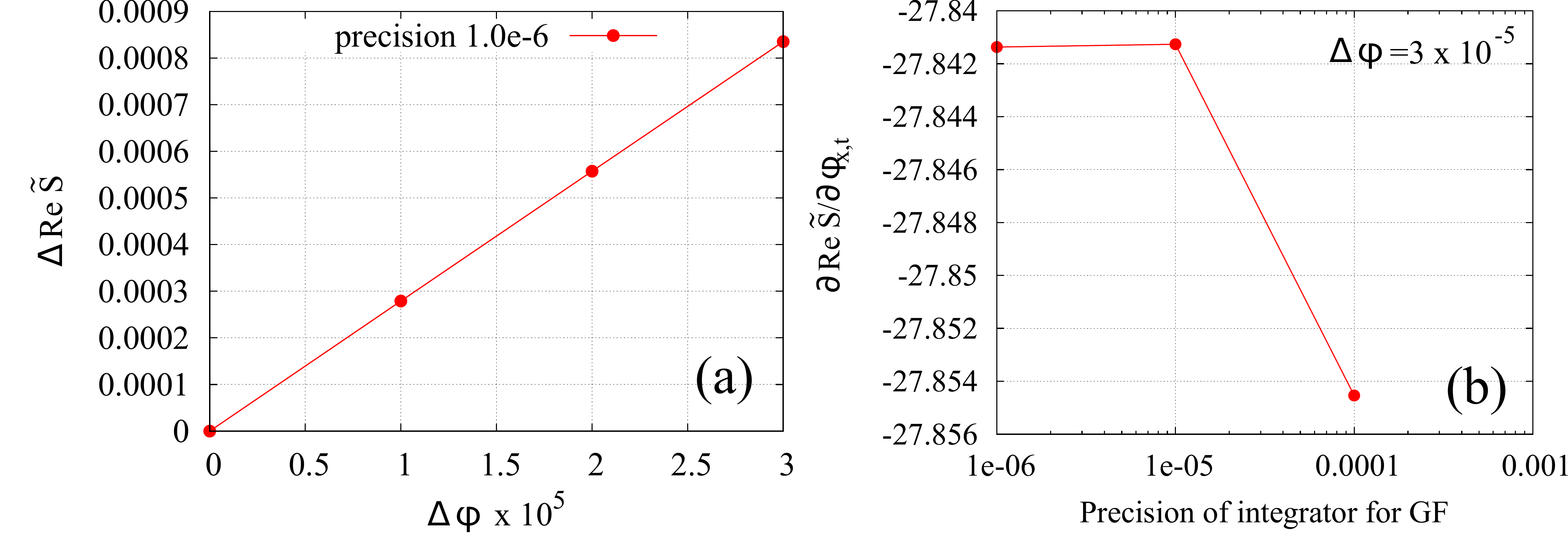}
        \caption{Technical plots demonstrating that the calculation of derivatives through finite differences is indeed reliable. (a) The difference of the real part of the action at the end of two flow procedures where the initial field configurations differ at a single site by a variable amount $\Delta \phi_0$. (b) Dependence of the derivative of the real part of the action computed after the flow on the precision of the integrator for the GF equations. As in (a), one starts from two initial field configurations differing at a single site ($\Delta \phi_0 = 3.0\times 10^{-5}$) and at the end of the flow one computes the derivative as a finite difference, $\Delta \mbox{Re} S/\Delta \phi_0$. Clearly, the derivative stabilizes once the precision is high enough. Usually we need around 20 steps in the GF procedure for typical flow lengths. These examples are shown for exactly the same setup which we are using in one of our HMC flow simulations: $\alpha=1.0$, $2\times2$ lattice with $N_{\tau}=256$ at $U=2.0 \kappa$, $\mu=\kappa$, $\beta =20.0$.}
        \label{fig:derivatives_HMC}
\end{figure}

\begin{figure}
        \centering
        \includegraphics[scale=0.18, angle=0]{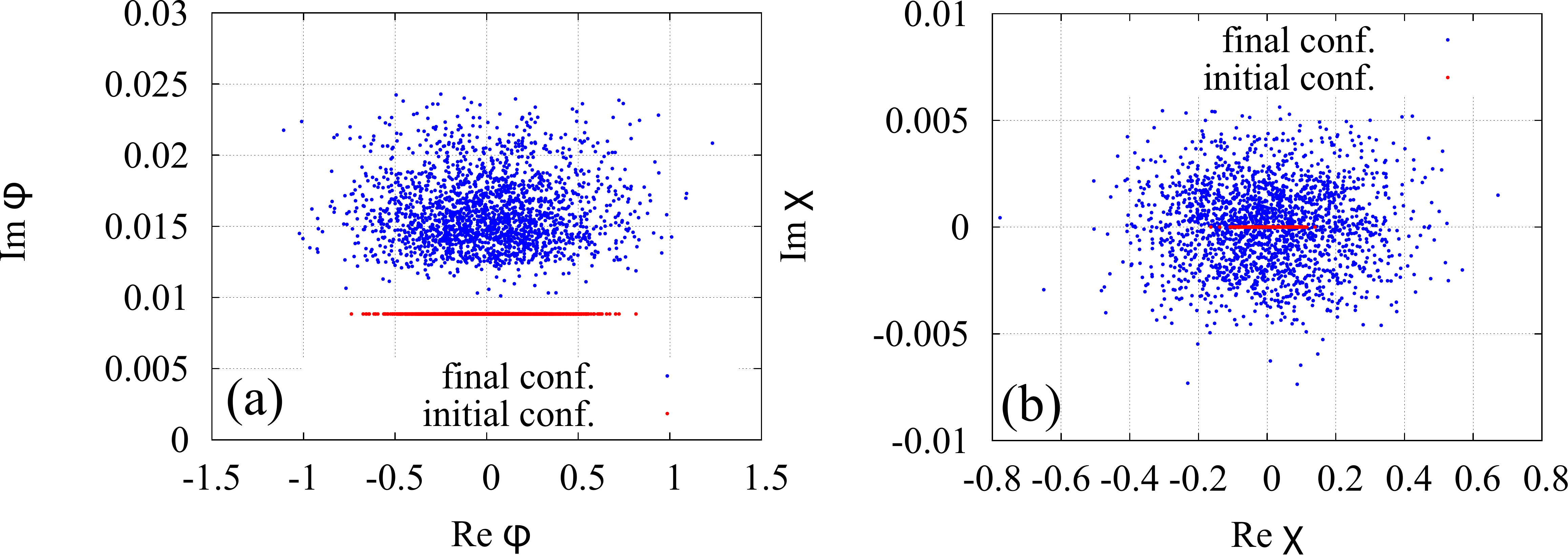}
        \caption{Example of configurations, generated in HMC with GF, for charge-coupled (a) and spin-coupled (b) auxiliary field. In both cases we show both the initial configuration located on the Gaussian thimble attached to the trivial vacuum saddle point and the configuration after GF. Parameters of the run: $\alpha=0.8$, $2\times2$ lattice with $N_{\tau}=256$ at $U=2.0 \kappa$, $\mu=\kappa$, $\beta =20.0$. Unfortunately, one can not deduce any simple relation which allows for a fit of the result of the GF with some local function:  $\mbox{Im} \phi_i = F (\mbox{Re} \phi_{i})$. }
        \label{fig:configurations_HMC}
\end{figure}

We use the following strategy to sample the partition function (\ref{eq:HMC_flow}): 1) The Jacobian is not taken into account in the Markov process employed to generate field configurations $\tilde \Phi$, and is left for the final reweighting;  2) The fields $\Phi$ are generated using HMC, according to the distribution $e^{-\mbox{Re} S[\tilde \Phi (\Phi + \Phi_0)]}$; 3) The fields $\tilde \Phi$ are computed through the gradient flow evolution. Several examples from the second stage of the process are shown in Fig.~\ref{fig:example_hmc_flow} for one particular site of the lattice. The second stage requires an additional comment. HMC employs global updates of the fields, using molecular dynamics (MD) governed by $H=\frac{1}{2} \sum_{i} p_{i}^2 + \mbox{Re} S[\tilde \Phi (\Phi)]$, where an artificial momentum, $p_{i}$, is introduced for each Hubbard field $\Phi_i$. In order to solve Hamilton's equations, we need to compute the derivative $\partial  \mbox{Re} S[\tilde \Phi (\Phi + \Phi_0)]  /   \partial \Phi_i$. We calculate this quantity by shifting the initial fields $\Phi_i \rightarrow \Phi_i+\Delta \Phi$ and solving the GF equations for each shift. Examples of such calculations are shown in Fig.~\ref{fig:derivatives_HMC}. These plots show that we can compute these derivatives to sufficient accuracy increasing the precision of the numerical integration for GF equations. 
These calculations also give us immediate access to all elements of the Jacobian. In practice, however, we need the $\det J$ only after the accept-reject step of the HMC procedure, while the trajectory typically consists of $\mathcal{O}(10^2)$ steps. Thus, the calculation of the Jacobian plays only a subdominant role in computational efforts. The overall scaling of the method is $C_1 \tilde{N}_{MD} \tilde{N}_{GF} N_s ^4 N_{\tau} ^2 + C_2 N_s ^3 N_{\tau} ^3$, where the first term corresponds to the HMC procedure used to generate field configurations and the second term accounts for the calculation of $\det J$ at the end of the trajectory. Here $\tilde{N}_{MD}$ refers to the number of steps in a MD trajectory which is typically $\mathcal{O}(10^2)$, $\tilde{N}_{GF}$ refers to the number of steps in the integrator for GF equations which is typically $O(10^1)$, and $C_1$ and $C_2$ denote volume-independent constants. In what follows, we will refer to this algorithm as HMC-GF.  Several examples of configurations of the $\tilde \Phi$ fields, generated with this algorithm, are show in Fig.~\ref{fig:configurations_HMC}.

\begin{figure}
        \centering
        \includegraphics[scale=0.18, angle=0]{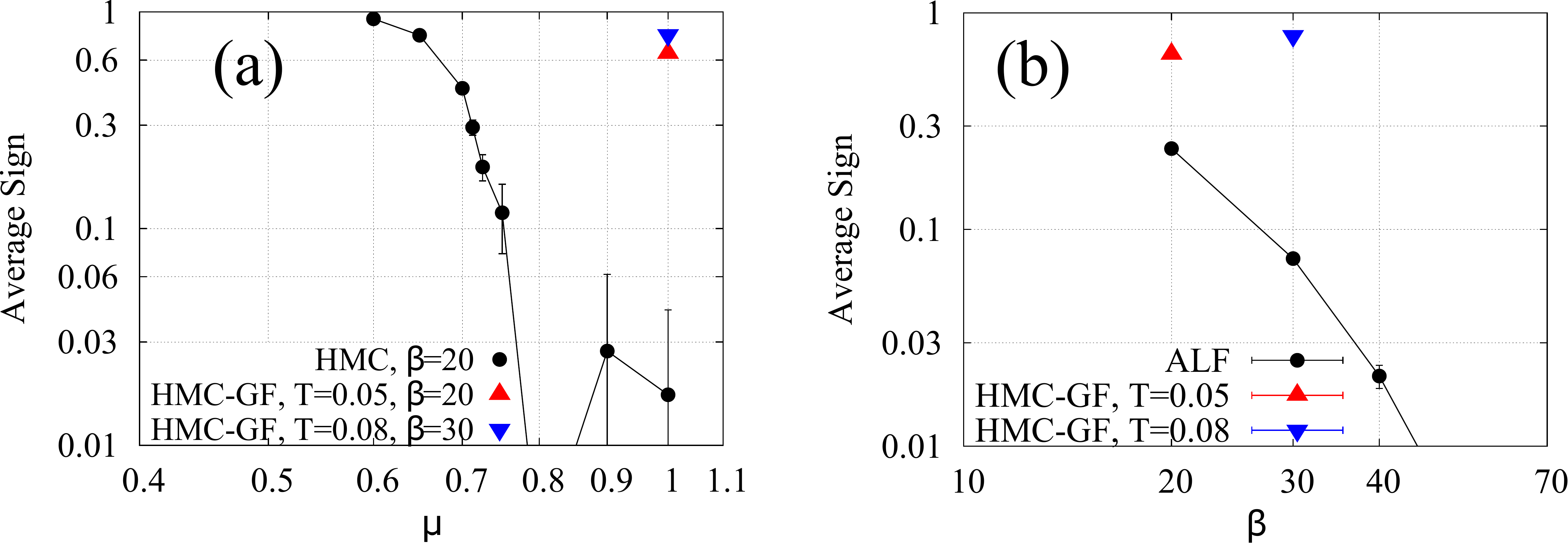}
        \caption{(a) Comparison of the sign problem in conventional HMC with real Hubbard fields and in HMC as a function of $\mu$. (b) Comparison of the sign problem in BSS-QMC and in HMC as a function of temperature at $\mu=\kappa$. Results are shown for a $2\times2$ lattice with $U=2.0 \kappa$, $N_\tau=256$. $\alpha=0.8$ for all HMC points.}
        \label{fig:sign_study}
\end{figure}

\begin{table}[t]
  \begin{tabular}{ | c | c | c |}
    \hline
                      & $\langle {\hat K} \rangle$ & $\langle \hat S^{(1)}_x \hat S^{(1)}_y \rangle$   \\  \hline  \hline
    ED                     & 19.5781             & -0.14624              \\ \hline
    BSS-QMC          & 19.587$\pm$0.002    & -0.1466$\pm$0.0008    \\ \hline
    HMC, $\alpha=1.0$ & 19.65$\pm$0.31      & -0.112$\pm$0.0069    \\ \hline
    HMC, $\alpha=0.8$ & 19.52$\pm$0.17      & -0.142$\pm$0.0062     \\ \hline
    \hline
  \end{tabular}
\caption{Comparison of observables for exact diagonalization, BSS-QMC (ALF) and two variants of HMC with gradient flow for a $2\times2$ lattice with $N_{\tau}=256$, $U=2.0 \kappa$ and $\mu=\kappa$.}
  \label{tab:observables}
\end{table}

\begin{table}[t]
  \begin{tabular}{ |  c | c | c | c |}
    \hline
                      & $\langle \cos \mbox{Im} S \rangle$  & $\langle \cos \arg J \rangle$ & $\langle \Sigma_G \rangle$  \\  \hline  \hline
    BSS-QMC          & 0.2363$\pm$0.0032 &                  & 0.2363$\pm$0.0032  \\ \hline
    HMC,  $\! \! \alpha \!\! = \! \! 1.0$ & 0.9627$\pm$0.0038 & 0.427$\pm$0.014  & 0.351$\pm$0.015    \\ \hline
    HMC,  $\! \! \alpha \! \!  = \! \! 0.8$ & 0.797$\pm$0.022   & 0.915$\pm$0.008  & 0.644$\pm$0.028    \\ \hline
    \hline
  \end{tabular}
\caption{Comparison of the sign problem for BSS-QMC (ALF) and two variants of HMC with gradient flow for a $2\times2$ lattice with $N_{\tau}=256$, $U=2.0 \kappa$ and $\mu=\kappa$.}
  \label{tab:sign}
\end{table}

The Jacobian is left for the final reweighting, and thus the observables are computed using the following expression
\begin{equation}
\label{eq:fin_observable}
\langle\mathcal{O}\rangle =\frac{\langle\mathcal{O} e^{i \, \mbox{\small{Im}} (-S +  \ln \det J) +   \mbox{\small{Re}} (\ln \det J) } \rangle }{ \langle  e^{i \, \mbox{\small{Im}} (-S +  \ln \det J) +   \mbox{\small{Re}} (\ln \det J) }\rangle}, 
\end{equation}
where the residual fluctuations of $\mbox{Im} S$ are also taken into account.
The brackets $\langle \rangle$ denote the averaging over configurations generated with HMC-GF. We also take into account symmetries of the action in order to further improve the ergodicity of our set of field configurations, generated with HMC-GF
\begin{eqnarray}
\label{eq:all_symmetries}
S(\phi_{x,\tau}, \chi_{x,\tau}) & = &  \bar S( - \bar \phi_{x,\tau}, - \bar \chi_{x,\tau}),
\nonumber \\ S(\phi_{x,\tau}, \chi_{x,\tau}) & = &  S(\phi_{x,\tau}, -\chi_{x,\tau}).
\end{eqnarray}
The following metrics are used to estimate the severity of the sign problem: $\langle \cos(\mbox{Im} S) \rangle$ and $\langle \cos(\mbox{Im}\ln \det J)\rangle$ for configurations and the Jacobian respectively, and the joint sign $\langle \Sigma_G \rangle = \langle \cos(\mbox{Im} (-S + \ln\det J))\rangle$. The first metric characterizes the part of the residual sign problem which stems from the fact that the sequence of shifts (\ref{eq:HMC_flow}) does not follow thimble exactly. The second metric characterizes the part of the residual sign problem which stems from the fluctuations of complex measure during integration over curved manifold in complex space. The last metric characterizes the entire residual sign problem. We also estimate the strength of the fluctuations of the Jacobian by computing $D_J$, the dispersion of $\mbox{Re} (\ln \det J)$.

The following choice is made for the parameters of the simulations: $2 \times 2$ lattice ($N_s=8$), $N_\tau=256$, $U=2 \kappa$, $\mu=\kappa$, $\beta=20$. This lattice is small enough to make a comparison with finite-temperature ED possible, but large enough to host non-trivial saddle points at large $\alpha$ (see Fig.~\ref{fig:charge_mu}\textcolor{red}{(c)}). Their form is only slightly different from the ones appearing at larger lattice sizes. These saddles also experience decay along the $\mbox{Re}\chi$ direction at $\alpha \approx 0.8$, similar to the $6 \times 6$ and $12 \times 12$ lattices studied above. Thus we can say that such a small lattice can in fact model the properties of the saddle points even at thermodynamic limit. On the other hand, we find that $N_\tau=256$ is large enough to probe both the low-temperature regime as well as the continuum limit in Euclidean time simultaneously. We further note that the state-of-the-art QMC algorithm for condensed matter systems, BSS-QMC, taken from the ALF package~\cite{ALF2017}, experiences exponential decay of the average sign at these parameters, even in the optimal regime where the discrete auxiliary field is coupled to spin. It is thus apparent that the sign problem is already strong in this regime. We have also probed two different values of $\alpha$: $\alpha=1.0$, so that only the charge-coupled field $\phi_{x,\tau}$ participates in the integral, and  $\alpha=0.8$ in order to probe the ``optimal regime'', where only the vacuum saddle point was detected.

Our results for the computed observables are displayed in the Tab.~\ref{tab:observables} while the study of the sign problem is summarized in Tab.~\ref{tab:sign}. We compute the kinetic energy, $\vev{\hat{{K}}}$, and the nearest-neighbor correlation function for the first component of spin $\vev{\hat{S}^{(1)}_x\hat{S}^{(1)}_y}$.   Results at $\alpha=1.0$ substantially deviate from ED, while at $\alpha=0.8$ the results of HMC calculation are in agreement with ED. This seems to imply that at $\alpha=1.0$ ergodicity issues indeed appear as there are several relevant thimbles and thus GF collides with zeros of the determinant. Unfortunately, HMC can not tunnel through the barrier separating two thimbles in such situations. At $\alpha=0.8$, however, ergodicity is restored. Moreover, we do not observe the growth of the fluctuations of the Jacobian, which should appear if GF approaches ``vertically'' oriented thimbles (Fig.~\ref{fig:schemes_hmc_flowa}). This tells us that the thimbles attached to the non-vacuum saddles indeed become irrelevant or they are bypassed by the integration manifold constructed by GF. In both cases, these non-vacuum saddles are effectively unimportant at $\alpha\approx 0.8$. Noting that there exists a striking resemblance between the saddle point structure on the $2 \times 2$ lattice and on larger lattices at large $\alpha$, table \ref{tab:observables} lends very strong evidence that at $\alpha\approx 0.8$, there exists an optimal regime with only one important thimble surviving in the full decomposition even for large spatial volumes. 

We also collected smaller statistics for the same lattice, but with $\beta=30$ with $N_\tau=384$ ($\alpha=0.8$). We increased the flow time from $\mathcal{T}=0.05$ for $\beta=20$ to  $\mathcal{T}=0.08$ for $\beta=30$ to keep the residual sign problem roughly the same ($\langle \Sigma_G \rangle=0.785 \pm 0.021$ for $\beta=30$ with $\mathcal{T}=0.08$ in comparison with $\langle \Sigma_G \rangle=0.644 \pm 0.028$ for $\beta=20$ with $\mathcal{T}=0.05$). As we have already mentioned introducing the Hubbard model, in this case, where only local interactions are present, we have the special possibility to perform calculations with BSS-QMC, using discrete auxiliary fields coupled with spin density. The sign problem in such set up seems to be several orders of magnitude smaller than in the approach with continuous auxiliary fields. Thus, in the Lefschetz thimbles formalism, where we can use only continuous fields, we start from a much stronger sign problem. However, despite this fact we can keep the average sign much higher than in BSS-QMC (see the Tab.~\ref{tab:sign} and the Fig.~\ref{fig:sign_study}\textcolor{red}{b}). Moreover, our calculations demonstrate another important property of the HMC-GF algorithm: despite the increased difficulty of the initial sign problem (average sign in BSS-QMC drops down in several times between $\beta=20$ and $\beta=30$), we can maintain the residual sign problem under control by increasing the flow time and moving the integration contour closer to the thimble: residual sign $\langle \Sigma_G \rangle$ even slightly increases for $\beta=30$ in HMC-GF. We expect similar behaviour for the dependence on lattice size $N_S$. The detailed study of the dependence on $N_S$ is left for upcoming papers, because we still need to improve our algorithms for constructions of the integration manifold for reasons listed below, and the estimates for the residual problem will likely change for modified algorithm.   

Unfortunately, larger average sign is still not enough to beat the better scaling of BSS-QMC: its scaling is given by $N_s^3 N_\tau$, which is substantially better than the  dominant term $N_s^4 N_\tau^2$ in the scaling of the HMC-GF. Thus, at these particular parameters, we can still use naive reweighting (generating $\sim 10^5$ configurations) to achieve error bars which are smaller than the ones computed in HMC-GF with $\sim 10^3$ configurations. 
The dispersion of the Jacobian is also noticeable but in general does not cause large problems. We simply need larger statistics to compensate for these additional fluctuations. We have determined $D_J=1.157$ for HMC-GF with $\alpha=1.0$ and $D_J=1.011$ for HMC-GF with $\alpha=0.8$. We noticed that the properties of the Jacobian become worse at $\beta=30$: $D_J=1.68$ and $\langle \cos \arg J \rangle = 0.823 \pm 0.018$ in this case (compare it with $\alpha=0.8$ case in Tab.~\ref{tab:sign}). Thus, at very low temperatures and, possibly, at larger system sizes, fluctuations of the Jacobian might become a problem. Right now we leave this potential problems for further study. 

As a final statement, we stress that from the point of view of lattice gauge theories, we should compare only with reweighting made with HMC running for continuous auxiliary fields, since there are no such cheap alternatives such as BSS-QMC for gauge theories or for the case of condensed matter systems with long-range interactions \cite{PhysRevB.90.085146}. The results of these benchmarks are shown in the Fig.~\ref{fig:sign_study}\textcolor{red}{(a)}. One can see that the improvement in comparison with plain HMC for continuous fields running inside $\mathbb{R}^N$ is quite huge. In fact, the van Hove singularity can not be reached with that method at all, as the average sign decays too quickly with increasing chemical potential. The plot essentially shows that even the present algorithm would allow us to go to the region completely inaccessible for standard HMC, thus it should already be the method of choice if we are simulating, for instance, the Hubbard-Coulomb model. Also, it suggests the feasibility of using the Lefschetz thimbles method for lattice gauge theories which suffer from the sign problem, since with HMC-GF in our studies of the Hubbard model we could reach the ratio $\mu/T=30$, at least for small lattices, while the sign problem typically becomes quite severe in lattice QCD at $\mu/T \sim 1$. However, the possibility of reaching such values of the parameters for QCD is at the moment purely speculative and should be checked explicitly in future dedicated simulations.

\section{\label{sec:Conclusion}Conclusion}
We proposed a set of new algorithms to study the properties of the thimbles decomposition for a lattice fermionic model approaching the thermodynamic limit. The method was tested for the case of the Hubbard model on the hexagonal lattice at various values of the interaction strength and chemical potential. Using this method, we could find the exact saddle points in the path integral formulation of the Hubbard model with different forms of the HS transformation, combining both the spin- and charge-coupled auxiliary fields with different weights. This represents the main physical result of the paper.
 At half-filling, in the case where the spin-coupled field dominates, we have observed a large number of different saddle points with an instanton-like structure. Away from half-filling, this regime develops a strong sign problem due to a large amount of thimbles with opposite phases which almost compensate each other. If the charge-coupled field is dominant, the structure of the saddle points is more regular, as they are built from two basic building blocks (localized field configurations) both at zero chemical potential and away from half-filling.  We have also observed an intermediate, ``optimal'' regime, where our method detected only one relevant saddle, even for a volume of  $12 \times 12\times 256$. These results show that the thimble decomposition for the Hubbard model strongly depends on the form of the HS transformation and that a solution to the sign problem using Lefschetz thimbles must properly take into account the saddle point structure. Consequently, the residual sign problem also depends strongly on the aforementioned decomposition of the fields.

In order to check these findings, we have performed several benchmark HMC simulations for lattices with $N_s=8$ and $N_\tau$ as large as $384$. In the ``optimal'' regime, we were able to reach an agreement with ED, which suggests that the ergodicity issues and the sign problem can be simultaneously weakened. We were also able to obtain an average sign which is much larger than that obtained with conventional QMC techniques. At the moment the thimbles algorithms is still too costly (mainly due to the aforementioned $N_s^4 N_\tau^2$ scaling with the system size), but the fact that the average sign can be kept higher than that of BSS-QMC gives us hope that we can construct a superior technique if we manage to decrease the computational costs associated with the integration over curved manifolds in complex space. Furthermore, the proposed set of algorithms can be helpful for the study of Lefshetz thimbles decomposition in strongly-interacting lattice models with fermions. The latter might be important for the better understanding of the corresponding physics through more accurate saddle point approximations of the path integral.

\begin{acknowledgments}
MU would like to thank Prof. F. Assaad for enlightening discussions. SZ acknowledges stimulating discussions with Prof. K. Orginos. 
MU is supported by the DFG under
grant AS120/15-1. CW is supported by the University of Kent, School of Physical Sciences. SZ acknowledges the support of the DFG Collaborative Research Centre SFB 1225 (ISOQUANT). This work was granted access to the HPC resources of CINES and IDRIS in France under the
allocation 52271 made by GENCI. We are grateful to these computing centers for their constant help. We are grateful to the UK Materials and Molecular Modelling Hub for computational resources, which is partially funded by EPSRC (EP/P020194/1). This work was also partially supported by the HPC Center of Champagne-Ardenne ROMEO.
\end{acknowledgments}

\bibliography{thimbles}

\end{document}